\documentclass[%
 article,
 jmp,%
 amsmath,amssymb,
preprint,50pt%
]{revtex4-1}

\usepackage{amsfonts}

\usepackage{bm}%
\usepackage[]{hyperref}
\usepackage{graphicx}
\usepackage{amsmath}
\usepackage{siunitx}
\usepackage{soul}

\makeatletter
\usepackage{rotating}
\usepackage{url}
\usepackage{anyfontsize}
\usepackage{blindtext}
\usepackage[utf8]{inputenc}
\usepackage{xr}
\usepackage{ccaption}

\usepackage{floatrow}
\usepackage[font= large,label font=bf,labelformat=simple]{subfig}

\usepackage[T1]{fontenc}
\usepackage{lipsum}
\usepackage[table]{xcolor} 
\usepackage{tabularx}

\begin{document}

\title{Spatially heterogeneous dynamics of cells in a growing tumor spheroid: Comparison between Theory and Experiments}
\author{Sumit Sinha$^1$, Abdul N Malmi-Kakkada$^2$, Xin Li$^2$, Himadri S. Samanta$^2$, D. Thirumalai$^2$}
\email{dave.thirumalai@gmail.com}
\affiliation{$^1$Department of Physics, University of Texas at Austin, Austin, TX 78712, USA.}
\affiliation{$^2$Department of Chemistry, University of Texas at Austin, Austin, TX 78712, USA.}

\begin{abstract}
Collective cell movement, characterized by multiple cells that are in contact for substantial periods of time and undergo correlated motion, plays a central role in cancer and embryogenesis. Recent imaging experiments have provided time-dependent traces of individual cells, thus providing an unprecedented picture of tumor spheroid growth. By using simulations of a minimal cell model, we analyze the experimental data that map the movement of cells in  fibrosarcoma tumor spheroid embedded in a collagen matrix. Both simulations and experiments show that cells in the core of the spheroid exhibit subdiffusive glassy dynamics (mean square displacement, $\Delta(t) \approx t^{\alpha}$ with $\alpha < 1$), whereas cells in the periphery exhibit superdiffusive motion, $\Delta(t) \approx t^{\alpha}$ with $\alpha>1)$. The motion of most of the cells near the periphery undergo  highly persistent and correlated directional motion due to cell doubling and apoptosis rates, thus explaining the observed superdiffusive behavior. The $\alpha$ values for cells in the core and periphery, extracted from simulations and experiments are in near quantitative agreement with each other, which is surprising given that no parameter in the model was used to fit the measurements.  The qualitatively different dynamics of cells in the core and periphery is captured by the fourth order susceptibility, introduced to characterize metastable states in glass forming systems. Analyses of the velocity autocorrelation of individual cells show remarkable spatial heterogeneity with no two cells exhibiting similar behavior. The prediction that $\alpha$ should depend on the location of the cells in the tumor is amenable to experimental test. The highly heterogeneous dynamics of cells in the tumor spheroid provides a plausible mechanism for the origin of intratumor heterogeneity.
 \end{abstract}


\maketitle
%



\section{Introduction}
Collective cell movement, involving a group of cells that are spatially adjacent and display coherent motion for long periods, control many processes such as embryogenesis,  tumorigenesis, and wound healing \cite{friedl2009collective,clark2015modes}. A number of factors, including intercellular interactions, cell division and apoptosis, regulate collective movement of cells \cite{friedl2009collective,ingber2003mechanobiology, guilak2009control, kumar2009mechanics, stetler1993tumor, tambe2011collective, clark2015modes,marchetti2013hydrodynamics}. In particular, the breakdown of  strict cellular homeostasis between cell division and apoptosis, which regulates tissue development and maintenance, could be a root cause of a number of cancers \cite{barres1992cell, weinberg2013biology}. Cancer metastasis is driven by migration of cells from the location of the primary tumor to secondary sites \cite{kumar2009mechanics}. There could be multiple  mechanisms underlying the migratory or invasive potential of cells \cite{friedl2009collective}.  However, elucidation of these mechanisms is difficult because of the highly coordinated and many body nature of their movement \cite{martin2013cancer, clark2015modes}. To decipher migration of tumor cells migrating into the surrounding matrix, experimentalists have studied individual cell dynamics in an evolving three-dimensional (3D) growing spheroid {\it{in vitro}} \cite{desoize2000multicellular, walenta2000metabolic, laurent2013multicellular,valencia2015collective} using a variety of imaging techniques. Typically, the spheroid is embedded in a collagen or extracellular matrix to mimic {\it{in vivo}} conditions \cite{valencia2015collective, richards20184d, martino2019wavelength,han2019cell,palamidessi2019unjamming}. These experiments have given direct glimpses of cell dynamics, which could be exploited to develop and test theoretical concepts in describing collective motion. 


A number of theoretical studies have reported different scenarios for the dynamics of cell motion. A pioneering  study \cite{ranft2010fluidization}, treating tissues using continuum elasticity theory supplemented by simulations,  showed that if the rates of cell division and apoptosis are equal in the steady state, then the long time dynamics is diffusive. In other words, the cell mean square displacement (MSD), which is measurable using imaging data, grows linearly with time. More recently, this prediction was confirmed using detailed computer simulations done in two dimensions, which established fluid-like behavior if the birth and apoptosis rates are balanced in the steady state \cite{matoz2017cell}. Thus, under these conditions the cells undergo fluid-like diffusion in the long time limit in contrast to the dynamics in confluent tissues in which cell division and apoptosis are prohibited {\cite{bi2016motility}.

A completely different scenario arises, due to non-equilibrium effects, if there is an imbalance between cell birth ($k_b$) and apoptosis ($k_a$) rates. Recently, we used simulations and theory to investigate the dynamics of cell in a growing three dimensional (3D) tumor in which $k_b$ and $k_a$ are unequal \cite{malmi2018cell,samanta2019origin}. To model cancer tumor growth, we used  $\frac{k_b}{k_a} \approx 20$. In addition to $k_a$ and $k_b$ and an appropriate short range cell-cell interaction, the cell dynamics depends on the microenvironment. In our model, the sensitivity to the microenvironment is expressed in terms of a dormancy criterion, determined by the local pressure ($p_i$) on the $i^{th}$ cell. If $p_i$  exceeds a critical pre-assigned value $p_c$, the cell becomes dormant until a time when $p_i < p_c$. The dormancy criterion serves as a mechanical feedback that limits the growth of the tissue \cite{shraiman2005mechanical,alessandri2013cellular,conger1983growth,puliafito2012collective}. We predicted that for this model  the dynamics of the cells is spatially heterogeneous. The cells exhibit highly persistent superdiffusive dynamics near the tumor periphery and subdiffusive glass-like dynamics in the core.
  

Here, we analyze imaging data of a growing 3D spheroid \cite{valencia2015collective} using our theoretical framework. We show that predictions of our model quantitatively capture the salient features noted in the experiment. There is a marked difference in the dynamics of cells in the core of the solid tumor compared to those at the periphery. Typically, the cells in the core exhibit glass-like dynamics characterized by subdiffusive motion (MSD increases sub linearly with time)  while those at the periphery undergo superdiffusive motion (MSD increases as a power law with an exponent that is greater than unity).  More importantly, the simulation results allow us to show that the dynamics of individual fibrosarcoma cells imaged in experiments also exhibit spatially  heterogeneous dynamics across the entire tumor. The extent of heterogeneity varies both spatially and temporally. This is also reflected in the length scale dependent fourth order susceptibility, introduced in the context of spin glasses and structural glasses, \cite{kirkpatrick1988comparison}, shows a peak for cells in the periphery whereas such a peak is absent for cells in the core of the tumor or appears at extremely long times. The velocity correlations for the cells in the core persist over multiple cell divisions. Remarkably, the massive heterogeneity is manifested in the velocity correlation of individual cells, which renders averages meaningless. This key finding has implication for observation of intratumor heterogeneity \cite{Alemendro13ARPathol, kirkpatrick2015colloquium,li2019share}. Comparison between experimental data and simulations, which were performed {\it without adjusting the model to fit the data}, clearly shows that collective cell motion exhibits high degree of spatial heterogeneity.  We argue that massive spatial and temporal heterogeneity could be a universal characteristic in collective cell dynamics, determined only by an imbalance in birth and apoptotic rates and mechanical feedback-driven tumor evolution provided the interactions between cells are short-ranged.

\section{Methods}
\textbf{Experimental Data:} We analyze the imaging data of the fibrosarcoma cells \cite{valencia2015collective} growing in a collagen matrix using simulations of a minimal model for tumor growth \cite{drasdo2005single, schaller2005multicellular}. It is important to describe briefly the relevant experimental details in order to appreciate that the model used in our simulations (explained in the next section) is adequate to provide insights into the experimental data. In the experiments, HT 1080 (human fibrosarcoma cell line) spheroid, with initial radius $174~\mu m$, was embedded in a three-dimensional collagen matrix to mimic {\it{in vivo}} conditions. The spheroids were grown over 7 days and the trajectories of a few cells (10 \% cells were labelled with enhanced Green Fluorescent protein) were tracked. The equatorial plane of the spheroid was imaged \textcolor{black}{at low magnification (10X) with a Nikon swept field microscope}, and the cells were tracked every 14 minutes for 8 hours (using Metamorph image recognition software) on day 3, 5 and 7 of spheroid growth. The number of cells tracked on day 3, 5 and 7 were 145, 157 and 150 respectively. The data comprises of the 2D projection of the 3D cell trajectory with the assumption that motion is isotropic in all directions. We were given  only the $x, y$ coordinates of individual cell trajectories.  For this work, we analyzed the data mainly for Day 7 unless mentioned explicitly. The fibrosarcoma cells divide every $\tau_{fib}= 21~hours$, which implies that the tumor spheroid grew over 8 cell cycle time . At the end of day 7, the radius of the tumor, $R_o$, was approximately $3~mm$.

\textbf{Simulations:} We briefly describe the simulation methods, which we developed previously \cite{malmi2018cell, malmi2019dual}.  We simulated a 3D growing tumor using an off-lattice model adopted from previous studies \cite{drasdo2005single, schaller2005multicellular}, where cells are considered as interacting soft deformable spherical particles which grow stochastically in time and divide into daughter cells on reaching a critical radius. The two body interactions between the cells are short ranged, consisting of two terms, a repulsion (elastic force) and attraction (adhesion). The magnitude of the elastic force ($F_{ij}^{el}$) between two cells of radii $R_i$ and $R_j$ is given by
\begin{equation}
F_{ij}^{el}=\frac{h_{ij}^{3/2}}{\frac{3}{4}(\frac{1-\nu_i^2}{E_i}+\frac{1-\nu_j^2}{E_j})\sqrt{\frac{1}{R_i(t)}+\frac{1}{R_i(t)}}},
\end{equation}
where $E_i$ and $\nu_i$ are the elastic modulus and Poisson ratio of the $i^{th}$ cell. The overlap distance between the two cells is denoted by $h_{ij}$. The adhesive interaction ($F_{ij}^{ad}$) is given by, 
\begin{equation}
F_{ij}^{ad}=A_{ij}f^{ad}\frac{1}{2}(c^{rec}_i c_j^{lig}+c^{lig}_i c_j^{rec}),
\end{equation}
where $A_{ij}$ is the overlap area between the two interacting cells and $f^{ad}$ determines the strength of adhesive bond ($c^{rec}_i=c^{lig}_i=1$). The net force (${\bf F}_i$) on the $i^{th}$ cell is the vectorial sum of elastic and adhesive forces that the neighboring cells exert on it. We performed over damped dynamics simulations  without thermal noise because the matrix viscosity is assumed to be large.  Therefore,  the equation of motion is taken to be $\dot{\textbf{r}}_i=\frac{{\bf F}_i}{\gamma}$,
where $\gamma$ is the friction term which models the matrix as a thick gel and ${\bf r}_i$ is the position of the $i^{th}$ cell.

In the simulations the cells grow stochastically and divide on reaching a critical radius. The growth of the $i^{th}$ cell is dependent on the microenvironment, which is determined by the pressure $p_i$ due neighboring cells.  If $p_i$ is smaller than a pre-assigned critical value, $p_c$, the cell grows in size. However, if $p_i > p_c$, the $i^{th}$ cell becomes dormant. The cell can switch between the dormant and growth mode depending on the ratio of $\frac{p_i}{p_c}$. The cell volume grows stochastically in time and it divides into two daughter cells (volume is conserved during cell division) on reaching a critical size. The growth of cell is controlled by cell cycle time ($\tau_{min}$), which was taken to be $15~hours$. Apoptosis can also take place in the simulations where a cell is randomly removed. The apoptosis rate is given by $k_a=10^{-6} s^{-1}$. Because $k_a << \frac{1}{\tau_{min}}$, we are simulating a growing system. Note that the cell cycle time in experiments ($\tau_{fib}=21~hours$) and simulations ($\tau_{min}=15~hours$) are comparable. It should be stressed that neither $\tau_{min}$ nor any other parameter was tweaked to obtain agreement with experimental data.

We initiated the simulations by placing 100 cells whose $x$, $y$, $z$ coordinates are chosen from a normal distribution with zero mean and standard deviation $20~\mu m$. The simulated tumor spheroid was evolved for $600,000~s$ or $11.1~\tau_{min}$. The trajectories of all the cells were recorded and analyzed in order to calculate dynamical observables that sheds light on heterogeneity.

\textcolor{black}{\textbf{Classification of Core and Periphery:} We arbitrarily classified cells as belonging to the core (periphery) if their distances from the center, $R_c$, is less (greater) than 1.5 mm (2 mm). Similarly, in simulations, cells with $R_c < 30 \mu m$ are classified as belonging to the core whereas cells with $R_c > 60 \mu m$, are assumed to be in the periphery. The only purpose of performing this analysis is to show that, on an average, the cell dynamics changes from the core to periphery.}

\section{Results}
\textbf{Sub-diffusive core and Super-diffusive periphery:} We first discuss our analysis of the experimental data. Given the small number of cells (150) imaged in the experiment \cite{valencia2015collective}, we divided the cells into two parts based on their distances from the center of the spheroid ($R_{c}$) (see the left inset in figure \ref{msd_expt}). We classified cells as belonging to the core (periphery) if their distances from the center, $R_c$, is less (greater) than $1.5~mm$ ($2~mm$). Since, imaging the cells in the core of a spheroid is technically difficult, there are fewer cells ($27$) in the core compared to the cells in the periphery ($100$). Mean Squared Displacement (MSD), $\Delta (t)$, is one of the metrics that can be readily evaluated from single particle trajectories \cite{giri2013arp2}. We evaluated \textcolor{black}{$\Delta(t-t_i,t_i)$}  using, 
\begin{equation}
\textcolor{black}{\Delta (t-t_i,t_i) = \frac{1}{N_{c}} \sum_{k=1}^{N_c} [{\bf r}_k(t) - {\bf r}_k(t_i)]^2},
\end{equation}
 where $N_c$ is the number of cells that belong to either the periphery or the core, and ${\bf r}_k(t)$ is the position of cell $k$ at time $t$. We denote  $t_i$ as the time when measurement of cell trajectory begins, and $t$ represents the time of the spheroid growth. In calculating $\Delta (t)$ we did not perform any time average because the spheroid is far from equilibrium, which could imply that the behavior of $\Delta(t)$ might depend on type of averaging performed \cite{metzler2014anomalous}. 

In general, we expect that $\Delta (t) \approx t^{\alpha}$. If $\alpha<1$, then the dynamics is subdiffusive, which could be suggestive of glass-like behavior. For a fluid-like motion $\alpha=1$. If $\alpha$ exceeds unity then the dynamics would be superdiffusive. Figure \ref{msd_expt} shows that  cells in the tumor core undergo subdiffusive dynamics with $\alpha=0.66$. In contrast, fibrosarcoma cells in the peripheral region undergo superdiffusive dynamics with $\alpha=1.34$ (see the right inset of figure \ref{msd_expt} for fits to $log(\Delta(t))$ vs $log(t)$, the slope of which determines $\alpha$).

In order to understand the spatially heterogeneous anomalous diffusion in a growing spheroid (\ref{msd_expt}), we simulated a freely expanding cell colony in 3D using the methods described elsewhere \cite{malmi2018cell,drasdo2005single,schaller2005multicellular}. We divided the simulated tumor spheroid into core and periphery. In the simulated tumor, cells with $ R_{c} < 30~\mu m$ are classified as belonging to the core whereas cells with $R_{c} >60~\mu m$, are assumed to be in the periphery. There exists a substantial length scale difference in what we define as periphery and core in simulations as compared to experiments because the size of the spheroid is on the order of $mm$ in experiments whereas the simulated spheroid reaches sizes on the order $\approx 0.2~mm$. However,  the simulations capture the experimental findings  well.  For the two spatial regions, we calculated $\Delta (t)$ for cells as was done for the experiments (figure \ref{msd_expt}). Figure \ref{msd_sim} shows that in the limit $t>\tau_{min}$, the MSD for cells in the interior is subdiffusive with $\alpha=0.58$, whereas the cells at the periphery exhibit superdiffusive behavior with $\alpha=1.52$. The plot was generated by tracking cells which were present in the simulation (note that cells can undergo apoptosis in simulations) between initial time $t_i \approx \tau_{min}$ and the final time $t_f \approx 11.1 \tau_{min}$ where $\tau_{min} = 54,000~ s$. We averaged the calculations over 50 such simulations. The $\alpha$ values extracted from simulations are in near quantitative agreement with experiments, which is remarkable given that no parameter in the model was adjusted to describe the experiments. Both experiments and simulations show that the cells at the tumor core display glass-like behavior ($\alpha<1$) and those in the periphery undergo superdiffusive ($\alpha>1$). 

\textcolor{black}{\textbf{Spatial variations in the cell motilities:} }The quantitative agreement with experiments for $\alpha$ values allow us to use simulations to provide nuanced analyses of the cell trajectories. We sub-divided the simulated tumor spheroid into four layers arbitrarily, and calculated $\Delta(t)$ for cells in each layer. \textcolor{black}{The thickness of each layer is roughly $25 \mu m$. We performed statistical averages using cells which were in the simulations between time $\tau_{min}$ and $11\tau_{min}$. In figure \ref{msd_r_sim}, the time dependencies of $\Delta (t)$ are plotted for the 4 layers}. 
\textcolor{black}{The results in figure \ref{msd_r_sim} reveal  two interesting aspects of the nature of cell motility inside a growing spheroid. (a) The $\Delta(t)$ curves exhibit non-uniform curvature on the timescale ($\approx 10\tau_{min}$). Nevertheless, to illustrate the spatial variations in the cell motilities, we fit the $\Delta(t)$ curves by a power law by dividing the total time into two intervals. One spans $T_{w1}= 10^5s<t<2.5\cdot10^5s$ and the other covers, $T_{w2}=3\cdot10^5s<t<5.5\cdot10^5s$.  The fits, in both the time intervals, reveal an  enhancement in the cell motility as one marches from core to the periphery. The extracted  effective exponents ($\alpha_{eff}^{T_{w1}}$ and $\alpha_{eff}^{T_{w2}}$) show that the cell motility changes from being sub-diffusive to super-diffusive as the distance from the center of the tumor increases. (b) The exponent values  (figure \ref{msd_r_sim} caption) in a given layer decreases ($\alpha_{eff}^{T_{w2}} < \alpha_{eff}^{T_{w1}}$)  as time advances because a cell in the periphery at a given time becomes part of the core at a later time. Thus, the values of $\alpha_{eff}$ are themselves time dependent, and their utility is to merely illustrate qualitatively the nature of the dynamics of the cells in an evolving tumor.}


 \textcolor{black}{The calculated exponents $\alpha_{eff}^{T_{w1}}$ and $\alpha_{eff}^{T_{w2}}$ (obtained by fitting $\Delta(t) \sim t^{\alpha_{eff}}$ in the first and second time window of figure \ref{msd_r_sim}) are shown in figure \ref{exp_r}.  We scaled the x-axis by $R_o$, which is the approximate radius of tumor spheroid.} In  experiments $R_o$ is $\approx 3~mm$ and for simulations $R_o$ is $\approx 0.1~mm $. The prediction that the effective diffusion exponent varies spatially as the distance from the spheroid center increases can be tested in experiments if the number of imaged cells is increased. We believe that light sheet microscopy methods could be used to test our predictions that the dynamics would change continuously from being jammed to exhibiting super-diffusive behavior \cite{martino2019wavelength, pampaloni2013high}. 
 
\textbf{van Hove function is non-Gaussian:} The anomalous nature of diffusion of cells inside the tumor spheroid can be gleaned by calculating the van Hove function (\textcolor{black}{$P(\Delta x, \delta t)$}) which gives the distribution of $\Delta x$ obtained from, 
\begin{equation}
\Delta x_i =  x_i(t+\delta t) -  x_i(t),
\end{equation}
where $x_i (t)$ is the $x$ coordinate of $i^{th}$ cell at time $t$. Figure \ref{fig:vanhove} shows \textcolor{black}{$P(\Delta x, \delta t)$}, in which $\Delta x$ has been time and ensemble averaged, for cells tracked in experiments ($\delta t=28~mins$) and simulations ($\delta t = 10~mins$). If the cells exhibited liquid-like dynamics then \textcolor{black}{$P(\Delta x, \delta t)$} would be a Gaussian \cite{hansen1990theory}. This expectation is in sharp  contrast with the nature of cell movement seen in figure \ref{fig:vanhove}. For cells in both the regions, \textcolor{black}{$P(\Delta x, \delta t)$}'s exhibit a fat tail in the distributions deviating substantially from Gaussian behavior. However, cells near the periphery take longer jumps indicating the fast movement of cells compared to cells in the core. As discussed elsewhere this is a manifestation of dynamic heterogeneity \cite{barrat1990diffusion,thirumalai1993activated,berthier2011dynamical}.

\textbf{Superdiffusive exponent is invariant under time translation:}
A growing spheroid is a non-equilibrium system, which means that the values of any physical observable could depend in principle on the time of measurement. To test whether the MSD exponent $\alpha$ depends on when the observation time cell trajectories are measured, we calculated time averaged MSD, $\Delta (t_d)$, on Day 3, Day 5 and Day 7 of spheroid growth. Time averaged MSD is defined as, 
\begin{equation}
\Delta (t_d) = \frac{1}{N} \sum_{i=1}^{N} \langle [{\bf r}_i(t_d+t) - {\bf r}_i(t)]^2\rangle_t , 
\end{equation}
$N$ denotes the total number of cells tracked and $\langle ... \rangle_t$ refers to time average. Figure \ref{day_3_5_7_expt}, shows $\Delta (t_d)$ measurements on days 3, 5 and 7. To our surprise, the exponents are independent of time with $\alpha \approx 1.4 $. This might mean that the tumor cells have not aged on the experimental time scale.

To ascertain if our simulations are  in accord with the analysis of the experimental data,  we calculated the time averaged MSD during different time periods of spheroid growth. Figure \ref{tumor_pic} shows the snapshots of simulations at $t=3\tau_{min}$, $t=5\tau_{min}$, $t=7\tau_{min}$ and $t=11\tau_{min}$. In the simulations, we considered cell trajectories for three time periods as done in experiments. The three periods were $3\tau_{min} < t< 4\tau_{min}$, $5\tau_{min} < t< 6\tau_{min}$ and $7\tau_{min} < t< 8\tau_{min}$.  The averaging was performed over all the cells that were present during both the beginning and at the end of measurement. Figure \ref{day_3_5_7_sim} shows the behavior of time averaged MSD during the three time intervals. Our simulations show the same behavior as obtained in experiments, with $\alpha=1.37$ for measurements during the three time intervals.

\textbf{Self-Overlap Function and Fourth Order Susceptibility:}
The extent of spatially heterogeneous dynamics can be further quantified using the self overlap function ($\Omega(l,t_d)$) \cite{kang2013manifestation,abate2007topological}, 
\begin{equation}
\Omega (l,t_d) = \frac{1}{N_c}\sum_{i=1}^{N_c}\Omega_i(l,t_d),
\end{equation}
where $N_c$ is the number of cells in core or periphery of the tumor spheroid,  $t_d$ is the delay time, and $l$ is the characteristic length scale associated with the overlap function $\Omega(l,t_d)$. The overlap function for the $i^{th}$ cell is given by
\begin{equation}
\Omega_i(l,t_d)=\langle \Omega_i(l,t_d,t) \rangle _t,
\end{equation}
and
where $\langle .... \rangle_t$ is an average over time. We calculated $\Omega_i(l,t_d,t)$ using,
 \begin{equation}
 \Omega_i(l,t_d,t)= \Theta(l-|{\bf r}_i(t+t_d) - {\bf r}_i(t)|).
 \end{equation}
  The length $l$, serves as the cutoff distance for which the Heaviside function ($\Theta(l-|{\bf r}_i(t+t_d) - {\bf r}_i(t)|)$) is equal to unity (zero), if $l$ is greater (smaller) than $|{\bf r}_i(t+t_d) - {\bf r}_i(t)|$. Thus, $\Omega_i(l,t_d)$ measures degree of movement of cells in the time $t_d$. We first calculated $\Omega(l,t_d)$ using the experimental imaging data. Figure \ref{overlap_expt}, shows the difference in the decay of $\Omega(l,t_d)$ of the cells in the core ($R_{c}<1.5~mm$) and the periphery ($R_{c}>2~mm$). The value of $l$ was chosen as $100~\mu m$ because on this length scale the difference between the dynamics of the interior and periphery cells are vivid (see figure \ref{msd_expt}). Figure \ref{overlap_expt} shows the stark difference in the dynamics of cells in the core, which exhibit slow dynamics compared to cells near the tumor boundary, which is also reflected in figure \ref{msd_expt}. The plot of $\Omega(l,t_d)$ for cells in the periphery was fit to an exponential ($Ae^{-\frac{t_d}{\tau}}$), which yielded $\tau=0.3\tau_{fib}$ ($\tau_{fib}$ is the cell doubling time for fibrosarcoma cells). 

 We also calculated $\Omega(l,t_d)$ from simulations using $l=\frac{10}{3}~\mu m$ which is small compared to $l=100 ~\mu m$, due to difference in spheroid sizes. However, the length scale $l$, in both experiments and simulations satisfy the criterion $\big(\frac{l}{R_o}\big)_{E}=\big(\frac{l}{R_o}\big)_{S}$, where $R_o$ is the radius of the tumor, and the subscripts $E$ and $S$ referent to experiments and simulations, respectively.  With this criterion the experimental and simulation results could be compared on equal footing. As mentioned earlier, $R_o$ for experiments is $3~mm$ and for simulation is $0.1~mm$. Figure \ref{overlap_sim} shows the difference in the overlap function of cells in the interior ($R_{c}<30 \mu m$) and  the periphery region  ($R_{c} > 60 \mu m$) for the simulated tumor spheroid. The behavior of the overlap function calculated in simulations qualitatively matches with the experiments for the core cells.  The exponential fit for the decay of  $\Omega(l,t_d)$ ($Ae^{-\frac{t_d}{\tau}}$) yielded $\tau_p=0.6\tau_{min}$ and $\tau_c=3.2\tau_{min}$ for the cells in the core and periphery respectively . The decay time $\tau$, for the cells the periphery, obtained using simulations ($\tau_p=0.6\tau_{min}$) is in good agreement with the experiments ($\tau=0.3\tau_{fib}$). However, it is difficult to compare the behavior of $\Omega(l,t_d)$, for cells in the core between experiments and simulations because the cells were not imaged for sufficient time in the experiments (the $\Omega(l,t_d)$ curve does not decay substantially. For cells in the core, the relaxation time $\tau_c=3.2\tau_{min}$ (see figure \ref{overlap_sim} obtained from simulations), is six times longer than $\tau_p$ ($\frac{\tau_{c}}{\tau_{p}}\approx 6$). Hence, we hypothesize that the imaging needs to be performed at least six times longer than current observation time to observe the relaxation of the $\Omega(l,t_d)$ for cells in the core.

\textbf{Spatial variations in $\Omega (l,t_d)$:} We sub-divided the tumor spheroid into multiple layers and calculated $\Omega (l,t_d)$ for cells in a given layer. Figure \ref{scat_layer} shows the dependence of $\Omega(l, t_d)$ as a function of distance from the center of the spheroid. The cells in the inner most layer execute very slow glass-like dynamics compared to the outermost layer. 
In order to further distinguish between the slow and fast dynamics in different layers in the spheroid, we calculated the fourth order susceptibility ($\chi_4(l,t_d)$) \cite{kirkpatrick1988comparison},
\begin{equation}
 \chi_4(l,t_d) = N_c[\langle \Omega_i(l,t_d,t)^2 \rangle - \langle \Omega_i(l,t_d,t)\rangle^2].
 \end{equation}
 Figures \ref{xi_4_expt} and \ref{xi_4_2_layer_sim} show $\chi_4(l,t_d)$ as a function of $t_d$ for cells in the core and periphery  in experiments and simulations, respectively. Both figures show qualitatively identical behavior with cells in the periphery exhibiting a peak in $\chi_4(l,t_d)$. We should note that in simulations, $\chi_4(l,t_d)$ for cells in the core exhibits a peak, which is absent in the experiments. As explained earlier this is because the peak in $\chi_4(l,t_d)$, which usually occurs when $\Omega(l,t_{peak})=\frac{1}{e}$, for cells in the core would occur at longer time scales ($\frac{\tau_{c}}{\tau_{p}}\approx 6$). 
 
 To understand the behavior of $\chi_4(l,t_d)$ as a function of $R_c$, we sub-divided the simulated  tumor spheroid in several layers.  Figure \ref{chi4_sim} shows the behavior $\chi_4(l,t_d)$ for cells as a function of distance from the center of spheroid ($R_c$). We note two interesting aspects from the behavior of $\chi_4(l,t_d)$. Firstly, the position of peak in $\chi_4(l, t_d)$, which corresponds to the maximal heterogeneity in the movement on cells at length scale $l$, shifts to the right due to the slow dynamics as we approach the center of tumor spheroid. Secondly, the amplitude of the peak in $\chi_4(l,t_d)$, which corresponds to growing dynamical correlation length \cite{donati2002theory}, initially increases (see inset of  figure \ref{chi4_sim}) and then decreases as a function of distance from the spheroid. 

\textbf{Cells in the periphery undergo directed and highly persistent motion: }
The massively heterogeneous nature of cell motility within a single tumor spheroid can be highlighted using the time-dependent changes in the  trajectories of individual cells. We first analyzed the directionality of  individual cell movement as a function of the distance from the center of spheroid ($R_{c}$) by calculating the Straightness Index ($SI$) \cite{d2017contact},
\begin{equation} 
SI (R_c) =\frac{1}{N_r} \sum_{i=1}^{N_r} \frac{|{\bf r}_i(t_f)-{\bf r}_i(t_i)|}{\sum |\delta {\bf r}_i(t)|}.
\end{equation}
The numerator in the above equation is the magnitude of the net displacement of the $i^{th}$ cell between time $t_i$ and time $t_f$. The denominator is the total length of the trajectory of the $i^{th}$ cell, and $N_r$ is the number of cells between $R_c$ and $R_c+\delta R_c$. For experiments, $\frac{\delta R_c}{R_o}=0.14$ whereas $\frac{\delta R_c}{R_o}=0.1$ for simulations. If $SI$ is unity, then  the cells move along a perfectly straight trajectory. Figure \ref{straightness_expt}, which displays the straightness of trajectories of cells calculated using experimental data in a growing spheroid on day 7 during 8 hours of imaging, shows clearly that straightness of the trajectory increases as the distance of the cell from the spheroid core increases. We also evaluated $SI (R_c)$ (see figure \ref{straightness_sim}) in simulations using  $t_i=\tau_{min}$ and $t_f=11.1\tau_{min}$.   The behavior of $SI (R_c)$ agrees  well with the trends observed in experiments. The cells in the core (periphery) have $SI\rightarrow 0$ ($SI\rightarrow 1$).

\textbf{Massive spatially heterogeneous dynamics:} To further illustrate the difference in the directed motility of cells in the periphery and the core, we calculated the persistence of individual cell movements in both experiments and simulations. We defined persistence ($P(t_d)$) using the velocity of the cells as,
\begin{equation} 
\label{Ptd}
P(t_d)=\frac{1}{N_c}\sum_{i=1}^{N_c} \langle \hat{v}_i(t+t_d)\cdot \hat{v}_i(t)\rangle_{t}. 
\end{equation}
In Eq. \ref{Ptd}, $\hat{v_i}(t)$ is the unit velocity vector of the $i^{th}$ cell at time $t$, $N_c$ is the number of cells in the spheroid core or the periphery, $t_d$ is the delay time, and $\langle ...\rangle_t$ refers to time average. Figures \ref{vel_auto_expt} and \ref{vel_auto_sim} show the $P(t_d)$ curves as a function of $t_d$ calculated from experimental data and simulations, respectively. We calculated $P(t_d)$ from simulations using the cells that were present during the time interval $10 \tau_{min}$ and $11.1 \tau_{min}$. Cells on the periphery move in a highly persistent (directed motion with hardly any decay in $P(t_d)$ as $t_d$ changes) manner compared to cells in the interior. The results in these figures show dramatically that there are substantial cell-to-cell variations in $P(t_d)$ with {\it no two cells exhibiting similar behavior}. In particular, there is  widespread heterogeneity in trajectories of individual cells (see $P_{t_d}$ for individual cells which are denoted by thin lines). This finding is also reminiscent of glassy systems, characterized by large subsample to subsample fluctuations within a single large sample of a glass \cite{Thirumalai89PRA}. The results in figures \ref{vel_auto_expt} and \ref{vel_auto_sim} imply that averages, shown in dark colors, have no physical meaning, and could provide misleading information.   The massively spatially heterogeneous dynamics of individual cells during collective movement might be a plausible mechanism for the origin of intratumor heterogeneity \cite{Alemendro13ARPathol,kirkpatrick2015colloquium,li2019share}.


\section{Discussion}
We have used simulations of a minimal model \cite{malmi2018cell} to analyze the experimental results \cite{valencia2015collective}, where individual cell trajectories were monitored using fluorescent microscopy in a tumor spheroid embedded in a 3D collagen matrix. Remarkably, without adjusting any parameter in the model to obtain agreement with experiment, the exponents characterizing the mean square displacement of cells in the core and periphery are in quantitative agreement with values extracted from experimental data. This allowed us to dissect the remarkable spatial and temporal variations in the dynamics of the cells from the center to the periphery of the tumor. Both experiments and simulations unveil that in the peripheral region of the spheroid, cells execute highly persistent super-diffusive dynamics whereas the motion in the tumor interior is  sub-diffusive. 

The sluggish cell dynamics in the core is reminiscent of relaxation in supercooled liquids as they undergo a transition to a glassy state \cite{kirkpatrick2015colloquium}. Using concepts from glass transition theory, we showed that higher order susceptibility for cells near the tumor periphery in experiments, which are fully accounted for in the simulations, shows a peak at $t \approx$ 5.6 hours - the approximate time over which coherent motion occurs. A similar calculation for the interior jammed cells shows a peak that is likely to be present at much longer time scales. The difference in the fourth order susceptibility illustrates the spatial and temporal heterogeneity. A fuller analyses of the simulation results confirm that the dynamics is massively heterogeneous with substantial cell-to-cell variations. The dynamics of individual cells  varies greatly depending on their spatial locations in the tumor. We predict that the exponents associated with the mean square displacement should change continuously as function of cell distance from the center of the spheroid. This prediction, which already has partial support (see Figure \ref{exp_r}), could be further tested by imaging experiments that track a much larger number than is currently possible.  

The excellent agreement between simulations, which were not intended to model the specifics of the growth of fibrosarcoma tumor spheroid in a 3D college matrix,  and experiments  allows us to suggest generic mechanisms that govern the growth of spheroids. Besides the short-range cell-cell interactions the parameters that control tumor expansion in our simulations are the asymmetry  between cell birth ($k_b$)  and apoptosis rates ($k_a$), and a dormancy factor that is expressed in terms of a pressure threshold that a cell experiences. The imbalance ($k_b \gg k_a$) produces self-generated active forces \cite{doostmohammadi2015celebrating} that act in a directed manner on cells that are close to the periphery, facilitating their persistent  motion. Such forces in cells are related to myosin-based contractile stresses, which have been argued to be a major factor in the directed growth \cite{valencia2015collective}. Our previous study also suggested (see especially Figure 14 in \cite{malmi2018cell}) that there must be a high degree of correlation in the movement of neighboring cells at the tumor periphery. In other words, the superdiffusive behavior is a consequence of collective correlated motion of cells near the boundary.  \textcolor{black}{In  an  expanding  tumor,  there is an outward radial  stress,  arising from an imbalance between the rates of cell birth and apoptosis,  which renders the cells on the periphery to be superdiffusive.} Because these arguments are general,  we propose that global dynamics of a growing spheroid must exhibit the features (super-diffusive motion in the periphery, jamming in the interior, and high degree of spatial heterogeneity in the movement of individual cells).  Finally, it is likely that the non-equilibrium dynamics, arising due to $k_b \gg k_a$ may also be relevant in other situations such as, embryogenesis, and wound healing. 

\textcolor{black}{A  posteriori rationale for observing super-diffusive behavior is that there is a radial flow that thrusts the cells at the boundary outward. Although this is certainly correct, it should be noted that the force leading to the radial velocity is not explicitly in the model but is {\it self-generated by the birth and apoptotic processes} \cite{malmi2018cell}. Moreover, such a force, which is inherent to the physics of tumor growth in the model, has to be persistent in order to observe  super-diffusive behavior (i.e, act over several cell doubling times). Moreover, the biologically relevant parameters ($p_c, \frac{k_b}{k_a}$) could be chosen to entirely suppress the super-diffusive behavior even though the tumor expands. Thus, the dynamics in the model is a complex interplay between short range forces as well as the criterion for dormancy, and cell birth and apoptosis rates. It is worth emphasizing that the good agreement between our findings and the analysis of the experimental data implies that similar mechanism is operative in the collective movement of fibrosarcoma cells against the collagen matrix. This, perhaps, is the major surprise in this study.}

\textcolor{black}{We have captured quantitatively the spatially heterogeneous dynamics of cells in a growing tumor. Analysis of the experimental data, which provide the time traces of a small number of individual cells \cite{valencia2015collective}, reveal that core cells exhibit sub-diffusive dynamics ($\Delta (t)\sim t^{\alpha}$, where $ \alpha=0.66$) and those near the periphery undergo super-diffusive dynamics ($\Delta (t)\sim t^{\alpha}$, where $ \alpha=1.34$). Remarkably, {\it without adjusting any parameter}, we predict that cells in core (periphery) exhibit sub-diffusive (super-diffusive) dynamics with $\alpha=0.57(1.52)$. Comparison with experiments show that there is only one potential limitation. Due to  differences in the size of the simulated and experimental tumor, we had to choose different length scales while comparing the overlap function and fourth order susceptibility. Nevertheless, the qualitative insights obtained from our work provides a way to explore the dynamics of tumor evolution by varying the parameters that are most relevant biologically ($p_c$, $k_a$ and $k_b$). Using the velocity autocorrelation function we revealed the massive dynamical heterogeneity of cells in an expanding tumor which makes the notion of mean of less relevance. This cell to cell variation is an example of phenotypic heterogeneity and our work will be important in providing a mechanism of the origin and maintenance of  intratumor heterogeneity.}

\textbf{Acknowledgement:}
We are grateful to  Dr. Angela M. J. Valencia for providing the imaging data of fibrosarcoma cells moving in tumor spheroid \cite{valencia2015collective}. We are grateful to an anonymous referee for pertinent comments on our manuscript. This work was supported by National Science Foundation PHY 17-08128. Additional support was provided by the Collie-Welch Reagents Chair (F-0019).

\clearpage
\floatsetup[figure]{style=plain,subcapbesideposition=top}
\begin{figure}
	\sidesubfloat[]{\includegraphics[width=0.52\textwidth] {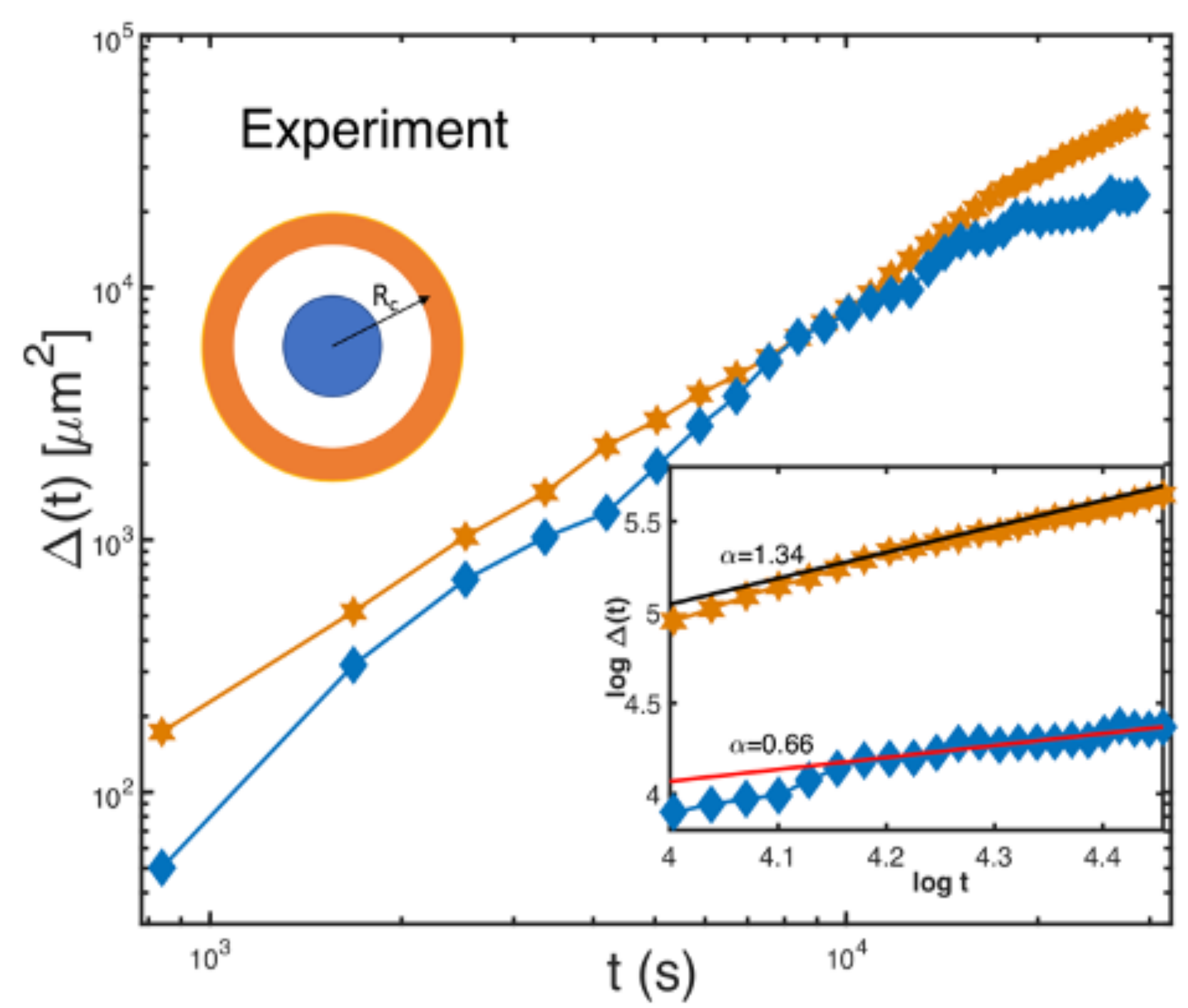} \label{msd_expt} }
	\sidesubfloat[]{\includegraphics[width=0.52\textwidth] {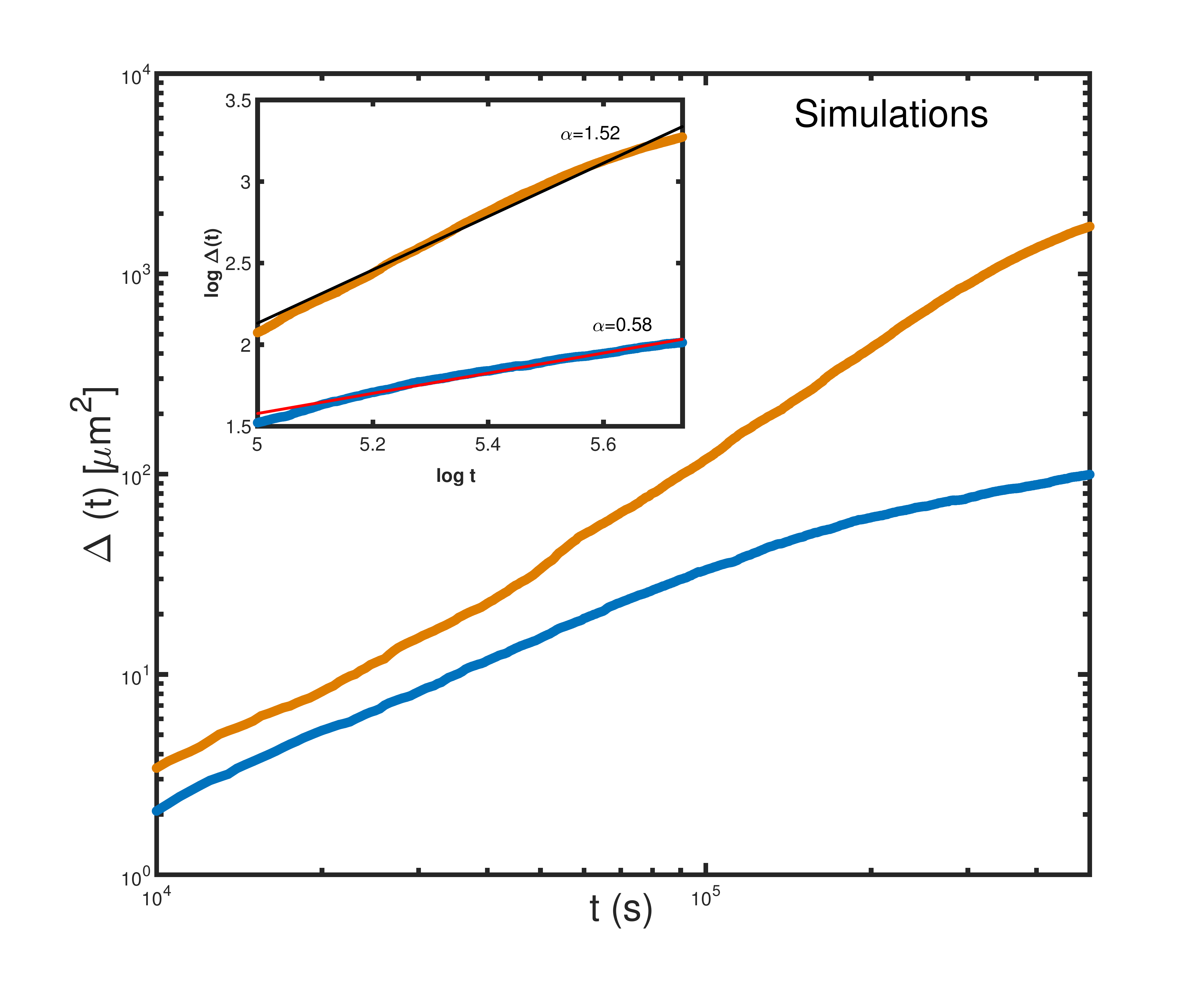} \label{msd_sim} }
	\par
	\sidesubfloat[]{\includegraphics[width=0.52\textwidth] {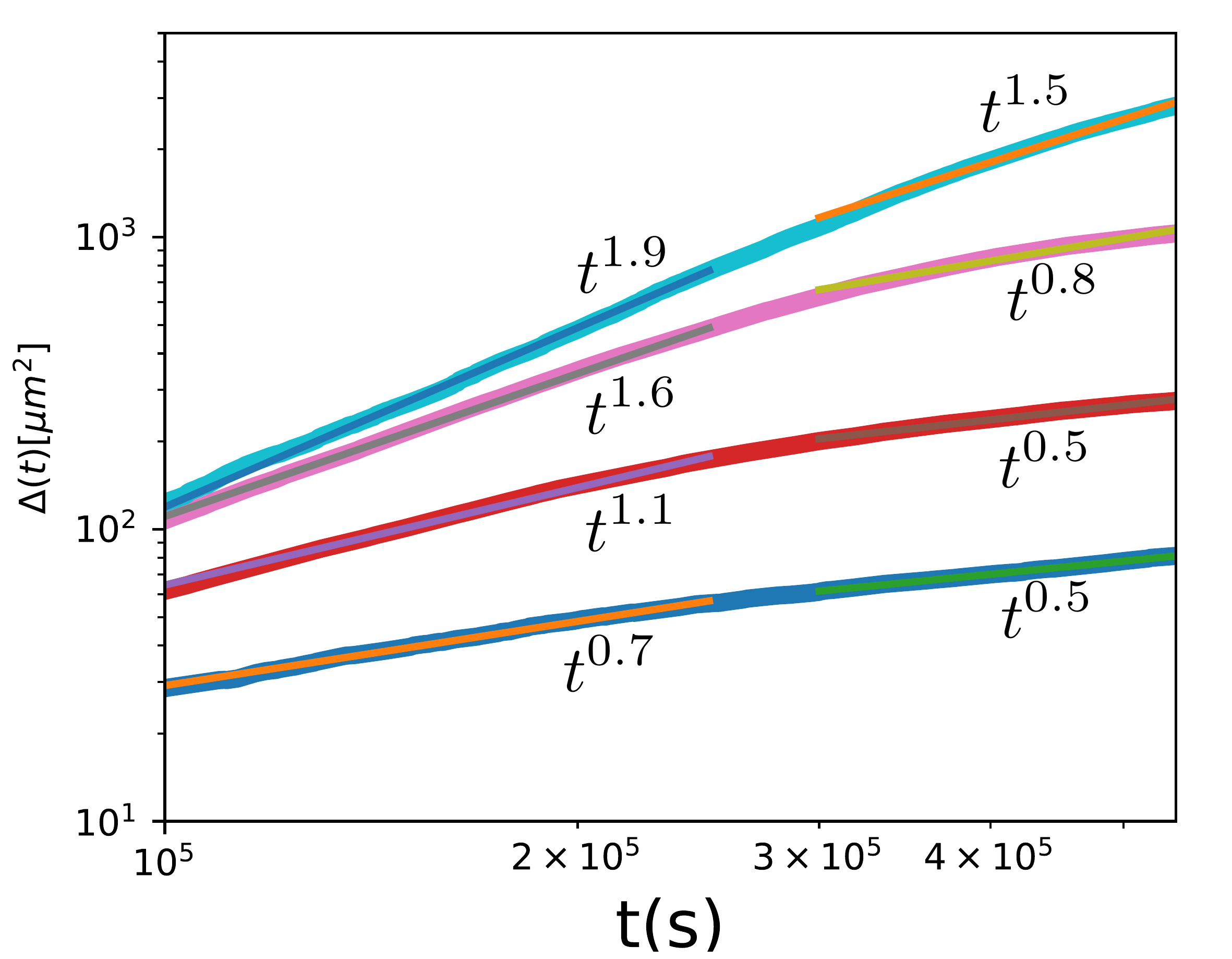} \label{msd_r_sim} }
	\sidesubfloat[]{\includegraphics[width=0.52\textwidth] {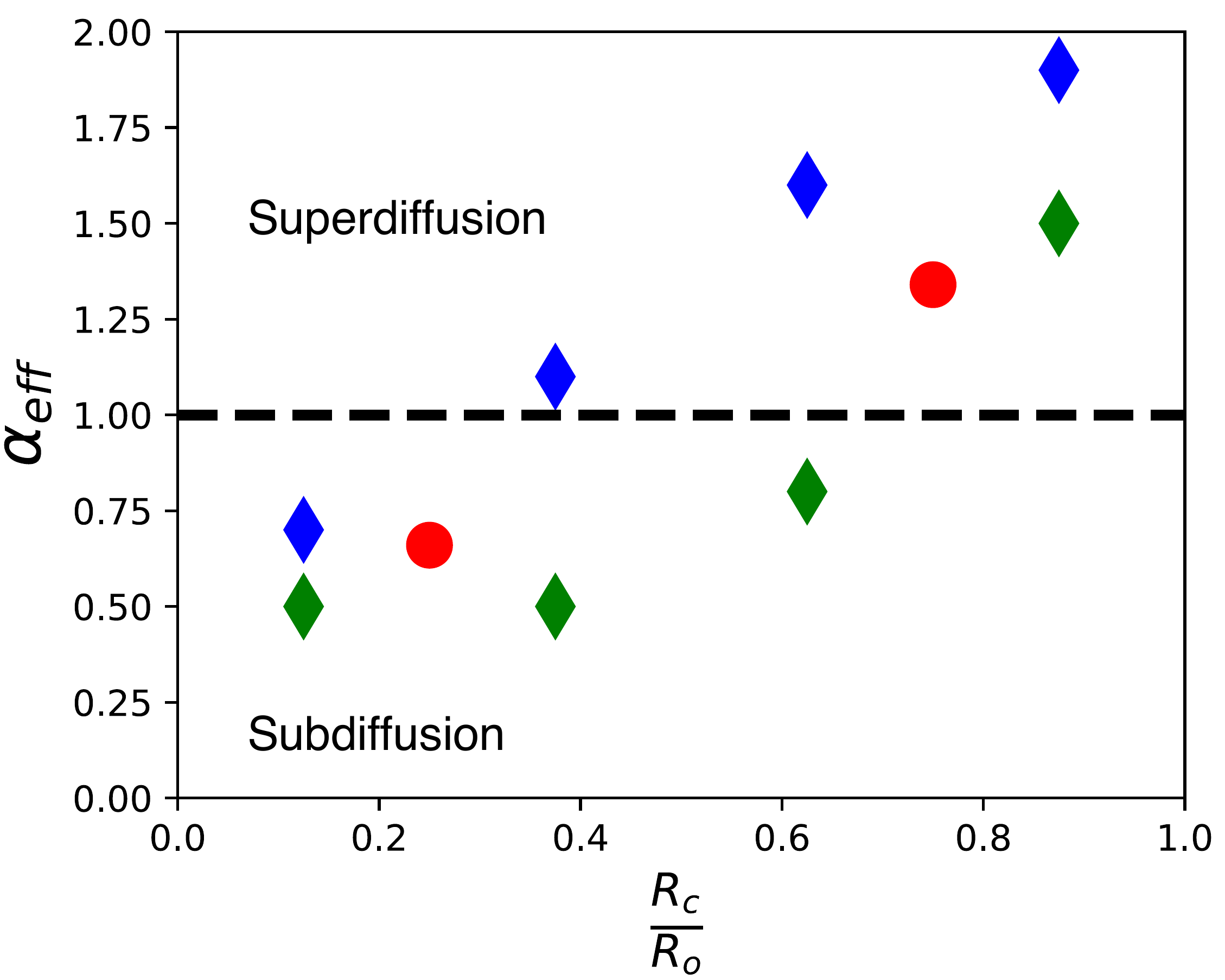} \label{exp_r} }
\caption{{\bf Spatial variation in dynamics.} \textbf{(a)} Mean Squared Displacement ($\Delta(t)$) as a function of time for experimentally tracked cells in the core (blue) and periphery (orange) of a growing spheroid. The measurements \cite{valencia2015collective} were performed on Day 7 of the growth. A schematic of the core and periphery in terms of $R_c$ is shown in the upper left. The blue line shows the MSD for cells in the core ($R_{c} < 1.5~mm$) and orange line depicts the MSD for cells in the periphery ($R_{c} >2~mm$). The inset shows the plot for $log(\Delta (t))$ vs ${log(t)}$ for cells in the core and periphery, where the periphery MSD has been multiplied by a factor of 10 for clarity. The slope of the curve $log(\Delta (t))$ vs $log(t)$, is the value of $\alpha$ in the equation $\Delta(t) \approx t^{\alpha}$. The black (red) line in the inset show the power law fit yielding $\alpha =1.34~(0.66)$. }
\label{msd_day_7}
\end{figure}
\clearpage
\begin{figure}
\contcaption{ \textbf{(b)} Same as (a) except $\Delta(t)$) has been calculated using simulations. The orange line shows MSD for periphery cells ($R_{c} > 60 \mu m$) whereas the blue line corresponding the MSD for cells in the core ($R_{c} < 30 \mu m$). The values of $\alpha$ are in black (red) for cells in the periphery (core). \textbf{(c)} MSD of cells in different layers in the growing spheroid calculated using simulations. \textcolor{black}{From bottom to top, the MSD curves are for cells whose distance from the center of the spheroid ($R_{c}$) are $0~\mu m < R_{c} < 25 ~\mu m$, $25~\mu m < R_{c} < 50 ~\mu m$, $50~\mu m < R_{c} < 75~\mu m$ and $75~\mu m < R_{c} < 100~\mu m$. The curves were fit by dividing time into two intervals: the first corresponds to $10^5 s<t<2.5\cdot10^5s$, and the second covers $3\cdot10^5 s<t<5.5\cdot10^5s$. The effective exponent values ($\alpha_{eff}$s) were calculated using,  $\Delta(t) \approx t^{\alpha_{eff}}$. The values of $\alpha_{eff}$  are given next to the curves .} \textbf{(d)} \textcolor{black}{Plot shows $\alpha_{eff}^{T_{w1}}$  for ${T_{w1}}$ and $\alpha_{eff}^{T_{w2}}$ for ${T_{w2}}$ as a function of $\frac{R_c}{R_o}$ ($R_o$ is the radius of the tumor defined in the text) for experiments (red disks) and simulations (blue diamond for the shorter time interval, and green diamond for the longer time interval). The dashed black line shows the line where $\alpha=1$ below (above) which denotes sub-diffusive (super-diffusive) motion.} }
\end{figure}
\clearpage

\floatsetup[figure]{style=plain,subcapbesideposition=top}
\begin{figure}

       {\includegraphics[width=0.65\textwidth] {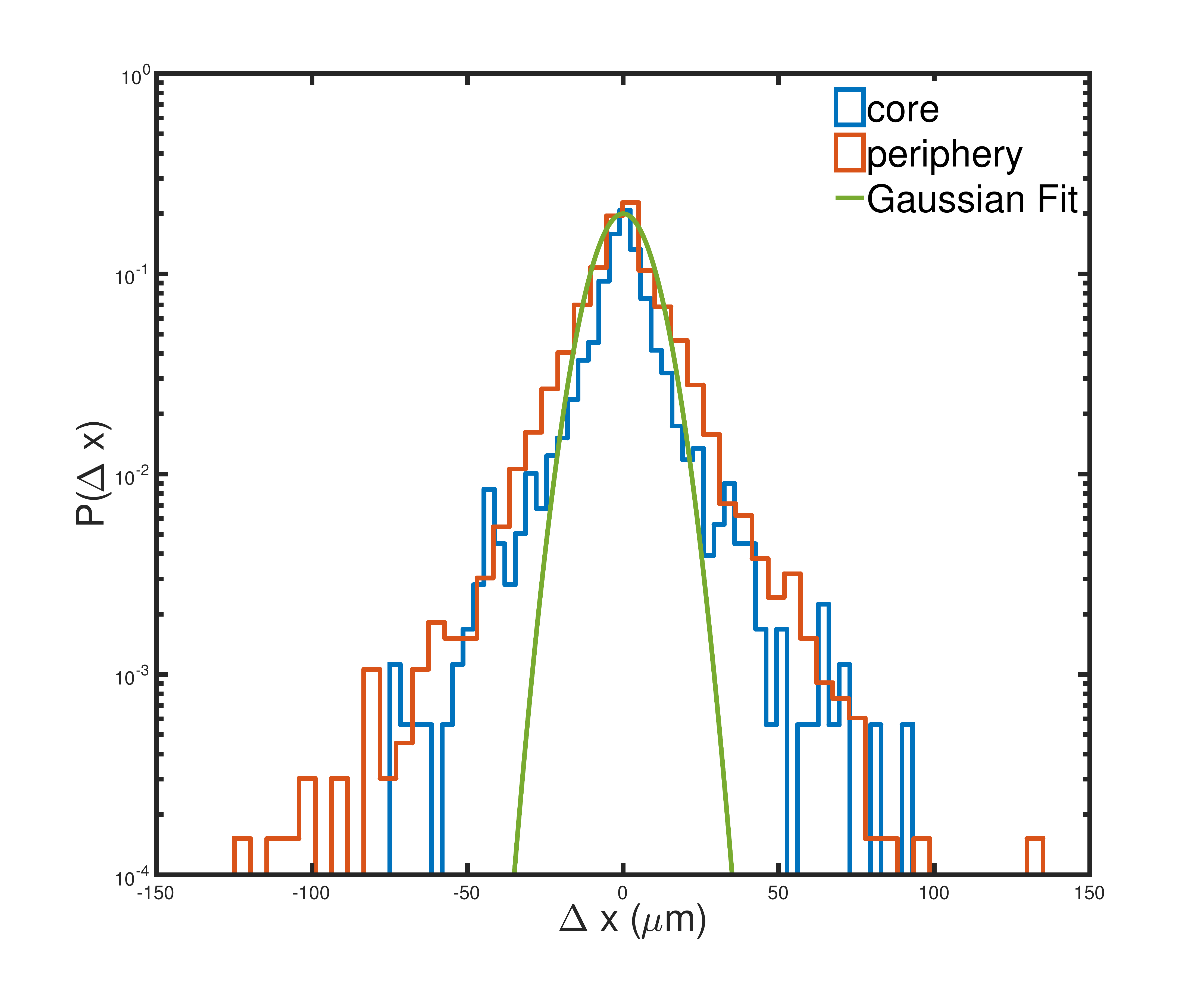} \label{vanhove_expt} }
       \par
       {\includegraphics[width=0.65\textwidth] {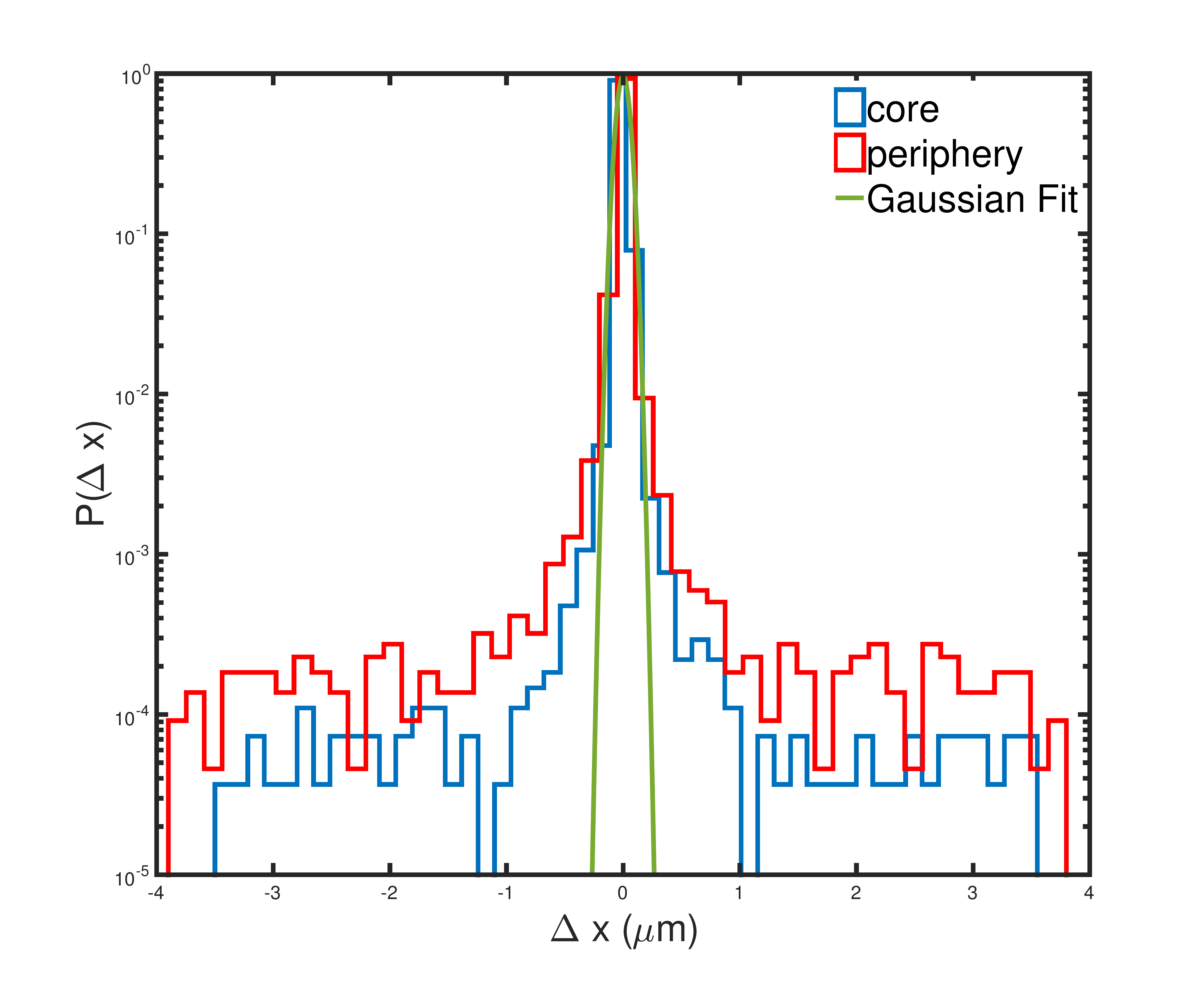} \label{vanhove_sim} }
\caption{{\bf Non-Gaussian behavior of cell displacements}. \textbf{(a)} van Hove function, ($P(\Delta x)$), for cells tracked in the experiments. The red (blue) line shows $P(\Delta x)$ for cells in the experiments. The green line is the Gaussian fit. \textbf{(b)} $P(\Delta x)$ for cells tracked in simulations where the red (blue) line are for cells in the periphery (core). The green line is the Gaussian fit. There is a striking similarity, except for the length scale, between simulations and experimental results. }
\label{fig:vanhove}
\end{figure}

\clearpage
\floatsetup[figure]{style=plain,subcapbesideposition=top}
\begin{figure}
        \sidesubfloat[]{\includegraphics[width=0.6\textwidth] {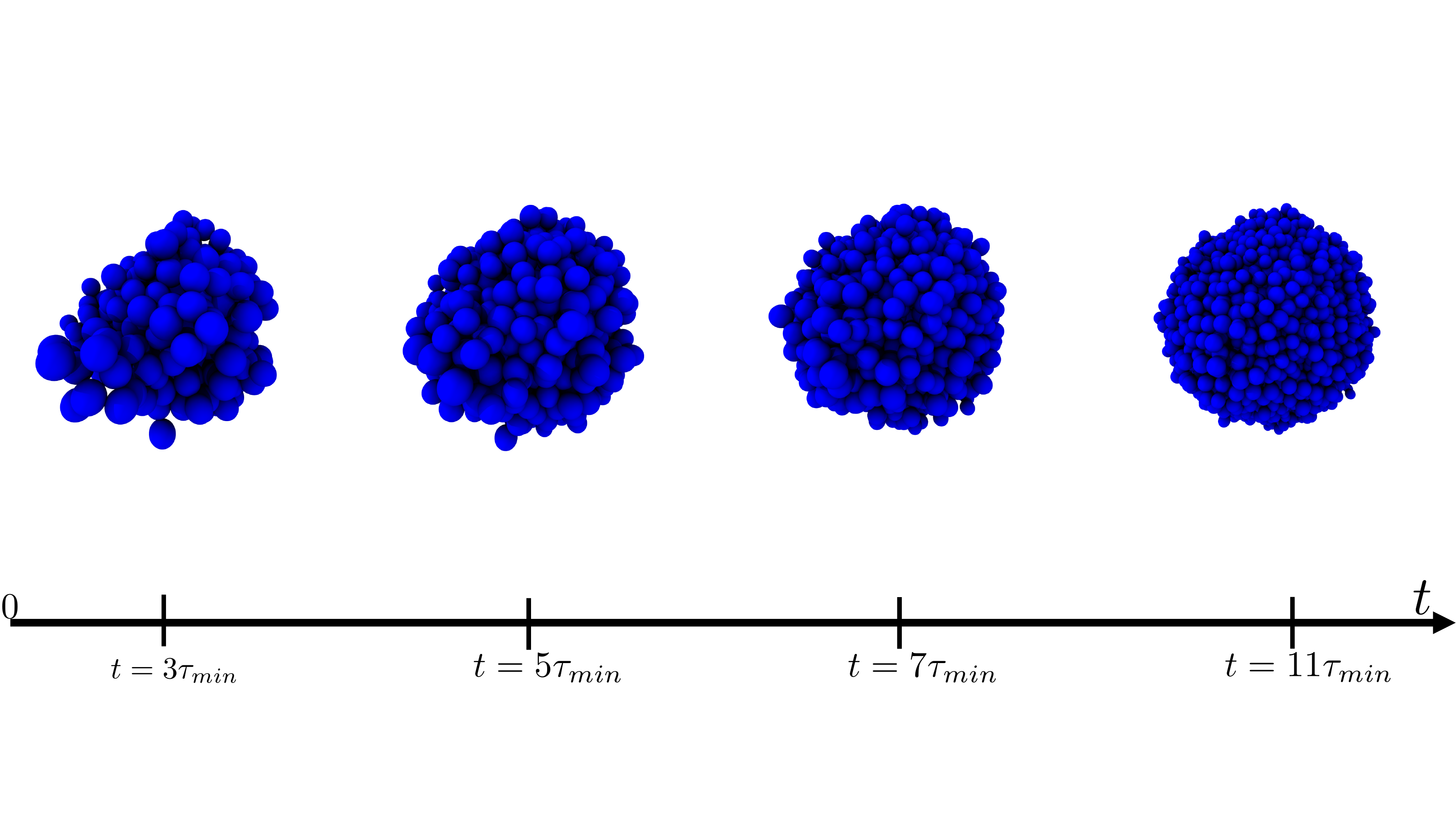} \label{tumor_pic} }
	\par
       \sidesubfloat[]{\includegraphics[width=0.55\textwidth] {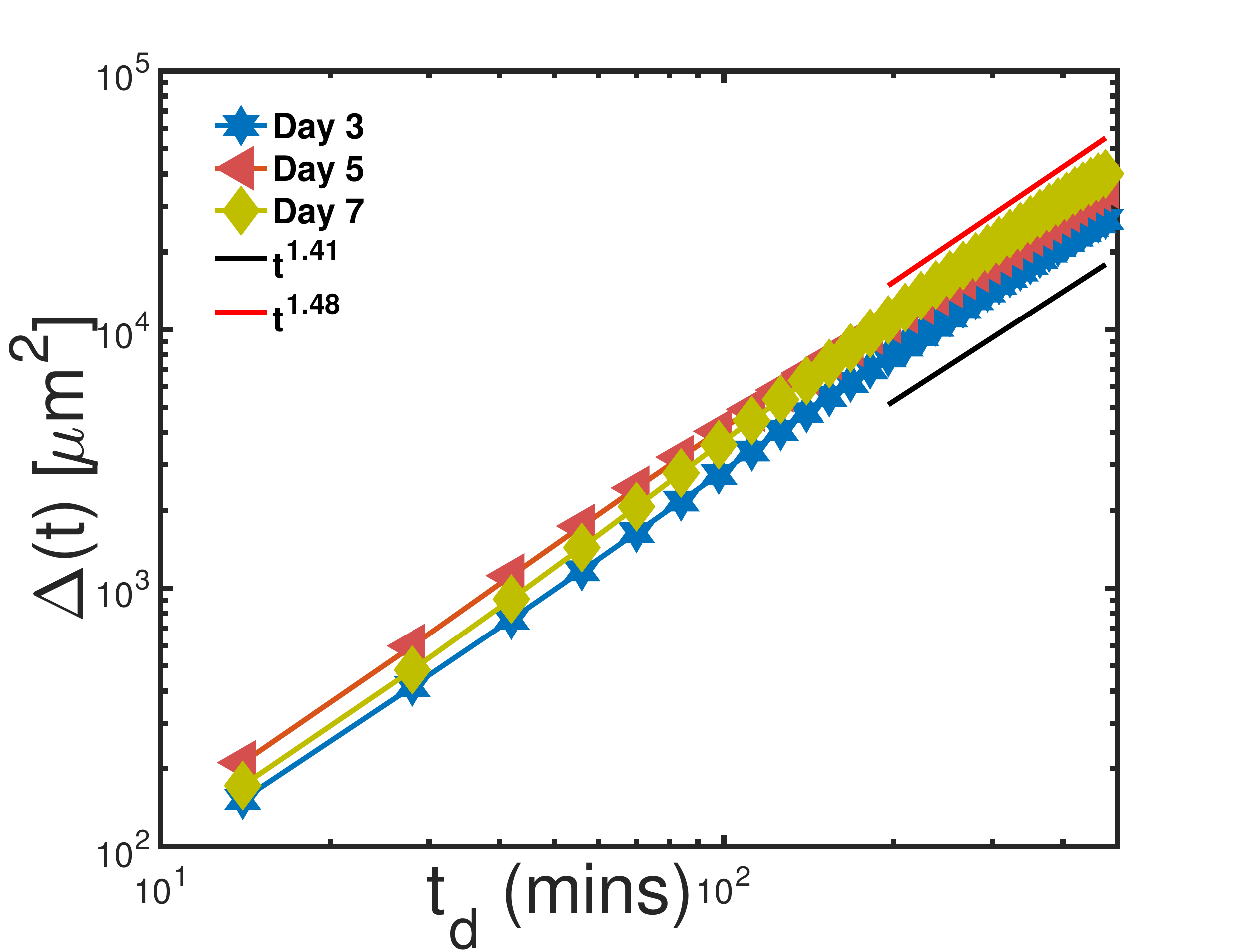} \label{day_3_5_7_expt} }
       \sidesubfloat[]{\includegraphics[width=0.55\textwidth] {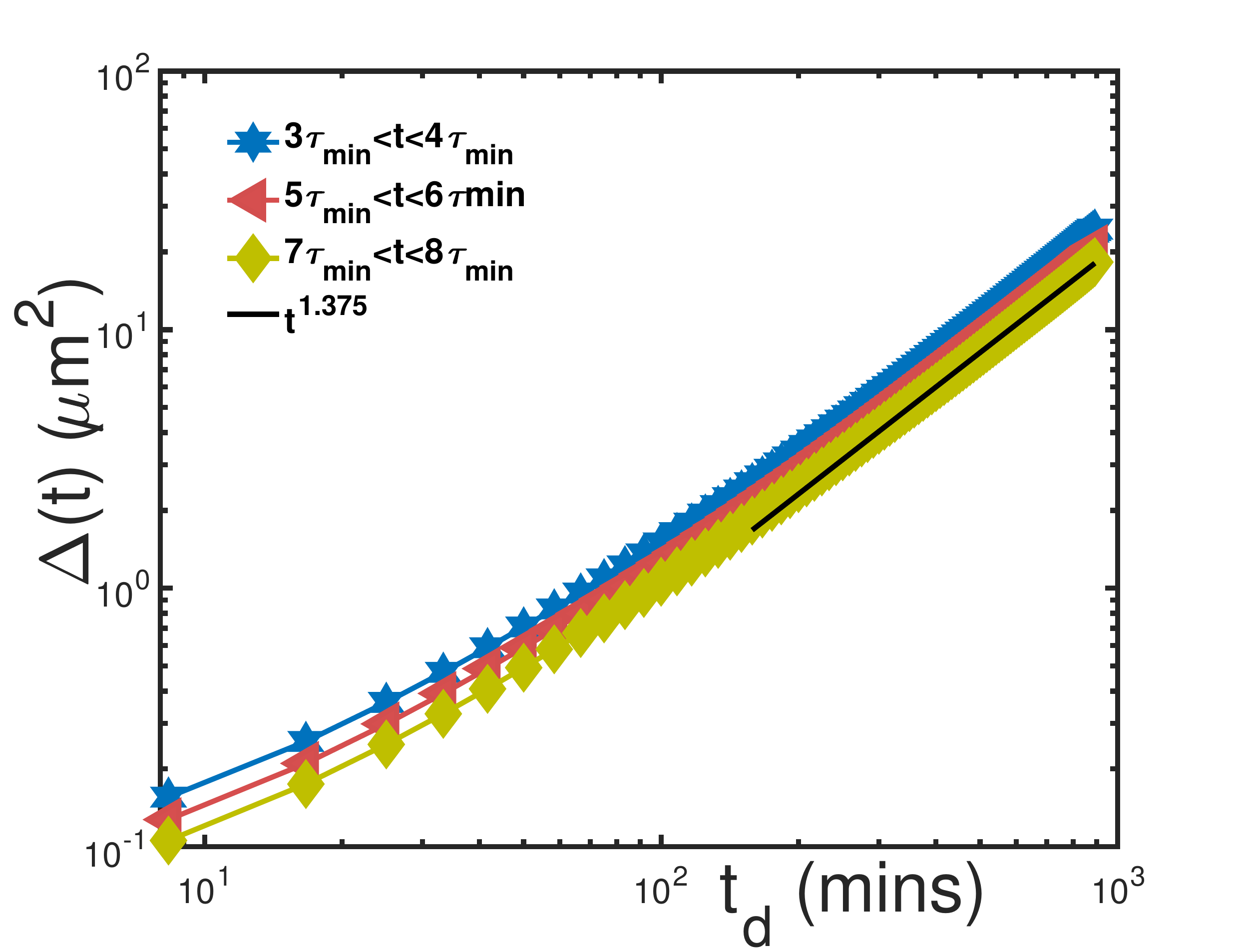} \label{day_3_5_7_sim} }
\caption{{\bf Long time MSD exponent is approximately independent of time in a growing spheroid}.\textbf{(a)} Snapshots from the simulations showing the growth of the tumor spheroid. The leftmost snapshot is at $t=3\tau_{min}$ ($\approx 500 $ cells), $t=5\tau_{min}$ ($\approx 1200$ cells), $t=7\tau_{min}$ ($\approx 2200$ cells) and $t=11\tau_{min}$ ($\approx 6000$ cells ). The black line below denotes the time axis with labels denoting the time of the snapshot.  \textbf{(b)} Time averaged $\Delta(t)$ for experimentally tracked cells on Day 3, 5 and 7. The blue line corresponds to Day 3, red line depicts Day 5 and green line shows Day 7. The black and red line show power law exponents of $1.41$ and $1.48$ respectively. \textbf{(c)} Time averaged $\Delta (t)$ of simulated cells for 3 observation times. The blue line corresponds to observation time of $3\tau_{min} < t< 4\tau_{min}$, red corresponds to observation time of $5\tau_{min} < t< 6\tau_{min}$ and green corresponds to observation time of $7\tau_{min} < t< 8\tau_{min}$. The black line corresponds to the power law exponent of $1.375$.   }
\label{fig:autocorrelation}
\end{figure}

\floatsetup[figure]{style=plain,subcapbesideposition=top}
\begin{figure}
	\sidesubfloat[]{\includegraphics[width=0.55\textwidth] {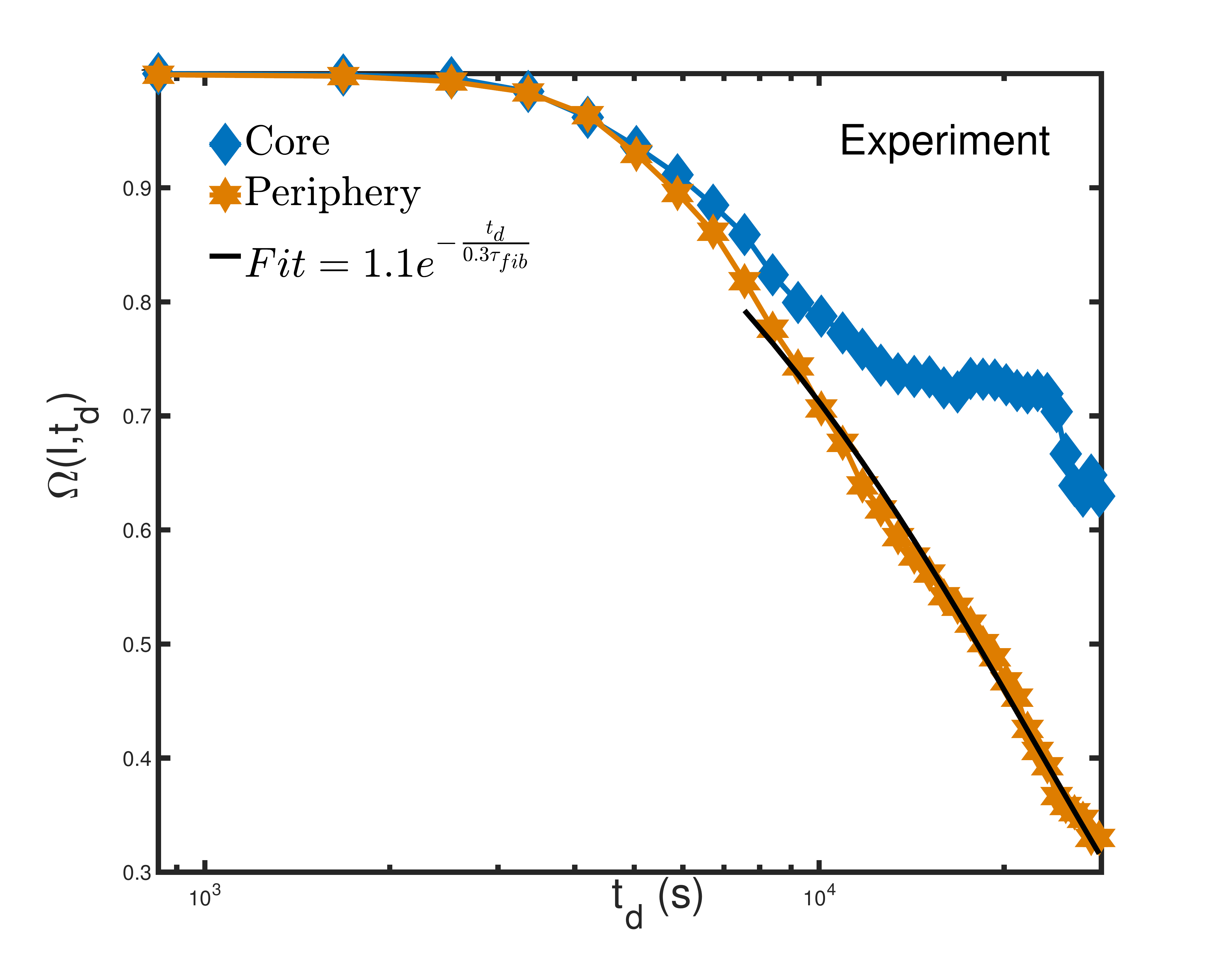} \label{overlap_expt} }
	\sidesubfloat[]{\includegraphics[width=0.55\textwidth] {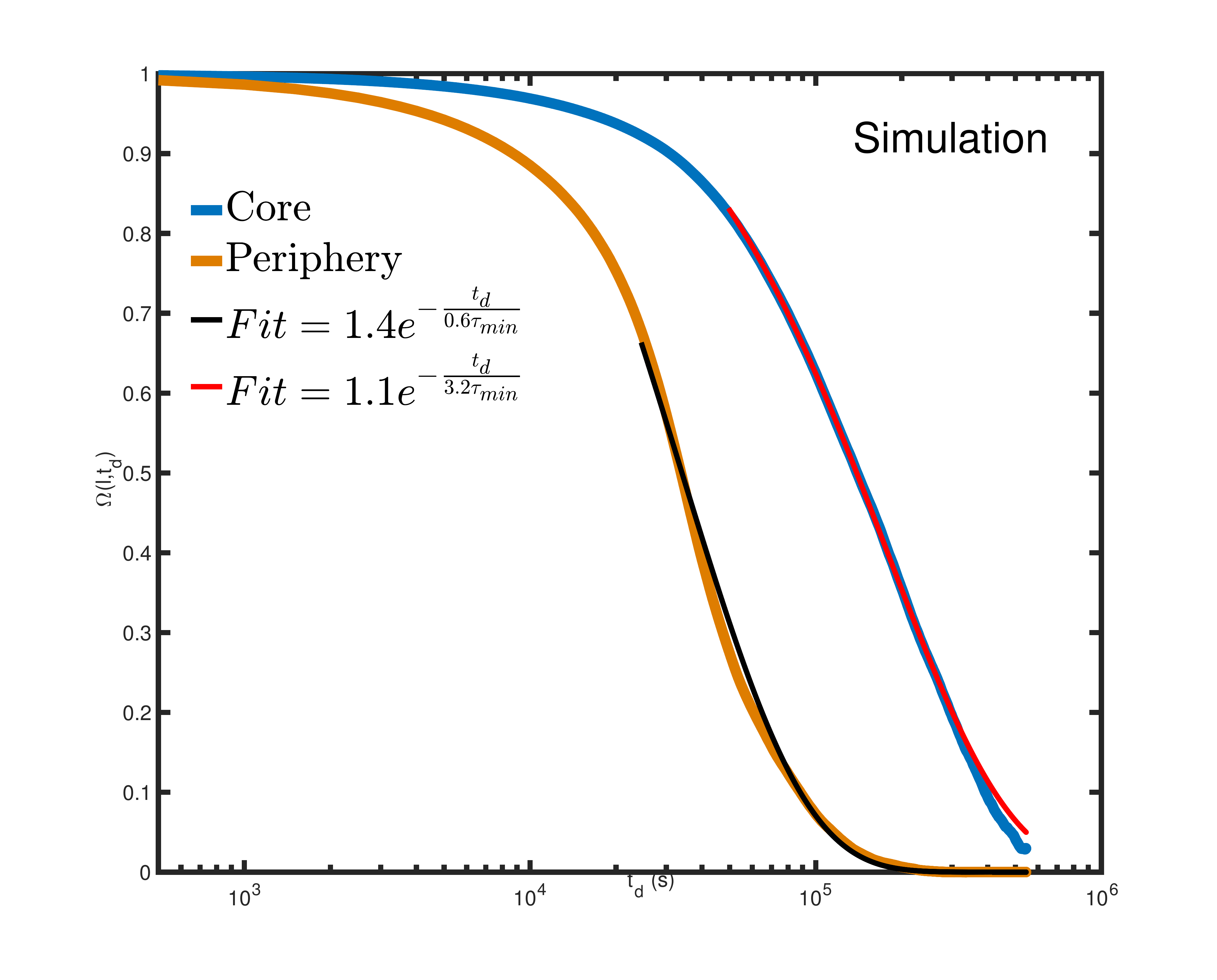} \label{overlap_sim} }
	\par
	\sidesubfloat[]{\includegraphics[width=0.55\textwidth] {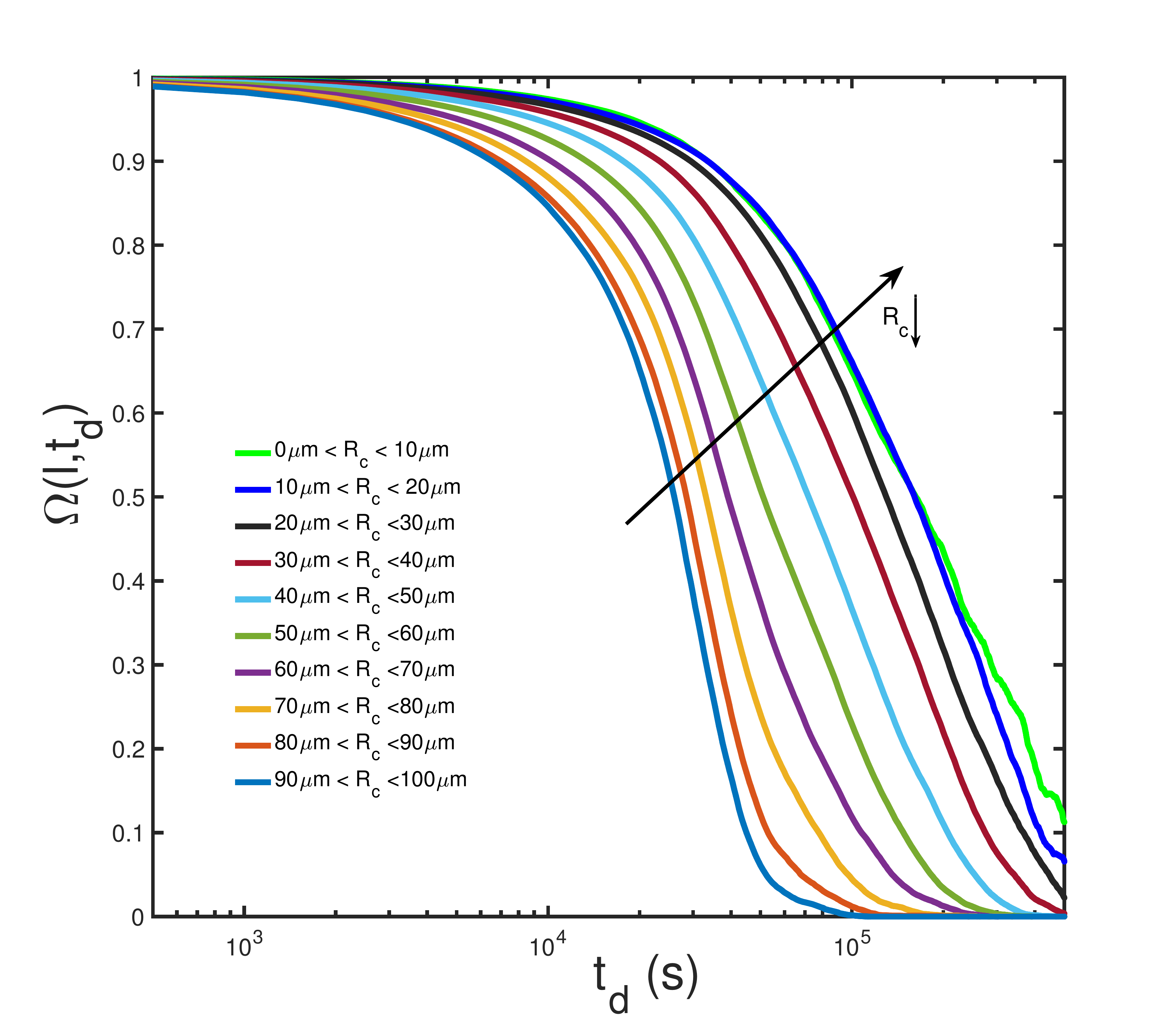} \label{scat_layer} }
\caption{\textbf{(a)} Self-Overlap function ($\Omega(l, t_d)$) for experimentally tracked cells as a function of delay time $t_d$ with $l = 100~\mu m $. The orange line shows the overlap function for cells near the periphery ($R_{c} > 2~mm$) and the blue line are for cells in the core ($R_{c} < 1.5~mm$). The black line is an exponential fit. \textbf{(b)} Self-Overlap function for simulated cells as a function of delay time $t_d$ with $l=\frac{10}{3}~\mu m$. The orange line shows $\Omega(l, t_d)$ for periphery cells ($R_{c} > 60 \mu m$) whereas the blue line corresponds to $\Omega(l, t_d)$ for cells in the core ($R_{c} < 30 \mu m$).\textbf{(c)} Time dependence of $\Omega(l,t_d)$ for cells in different layers of the spheroid. From top to bottom, $\Omega(l,t_d)$ curves are for cells whose distance from the center of spheroid ($R_{c}$) are $10(i-1)~\mu m < R_{c} < 10i ~\mu m$, for all $i = \{1,2,....,10\}$. The dashed arrow indicates the decreasing distance from the center of tumor spheroid along its direction ($R_c \downarrow$).}
\label{overlap_day_7}
\end{figure}

\clearpage
\floatsetup[figure]{style=plain,subcapbesideposition=top}
\begin{figure}
	\sidesubfloat[]{\includegraphics[width=0.55\textwidth] {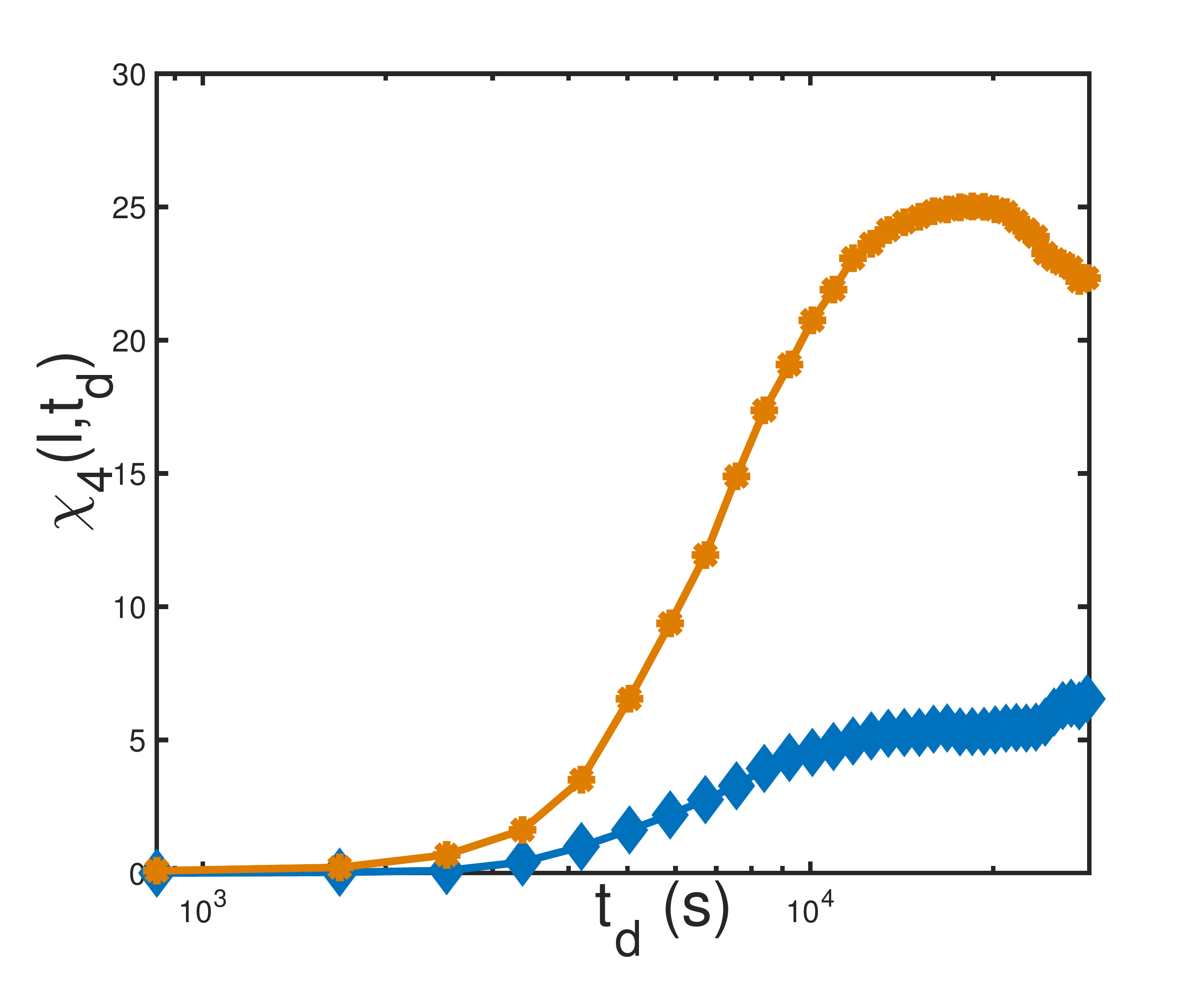} \label{xi_4_expt} }
	\sidesubfloat[]{\includegraphics[width=0.55\textwidth] {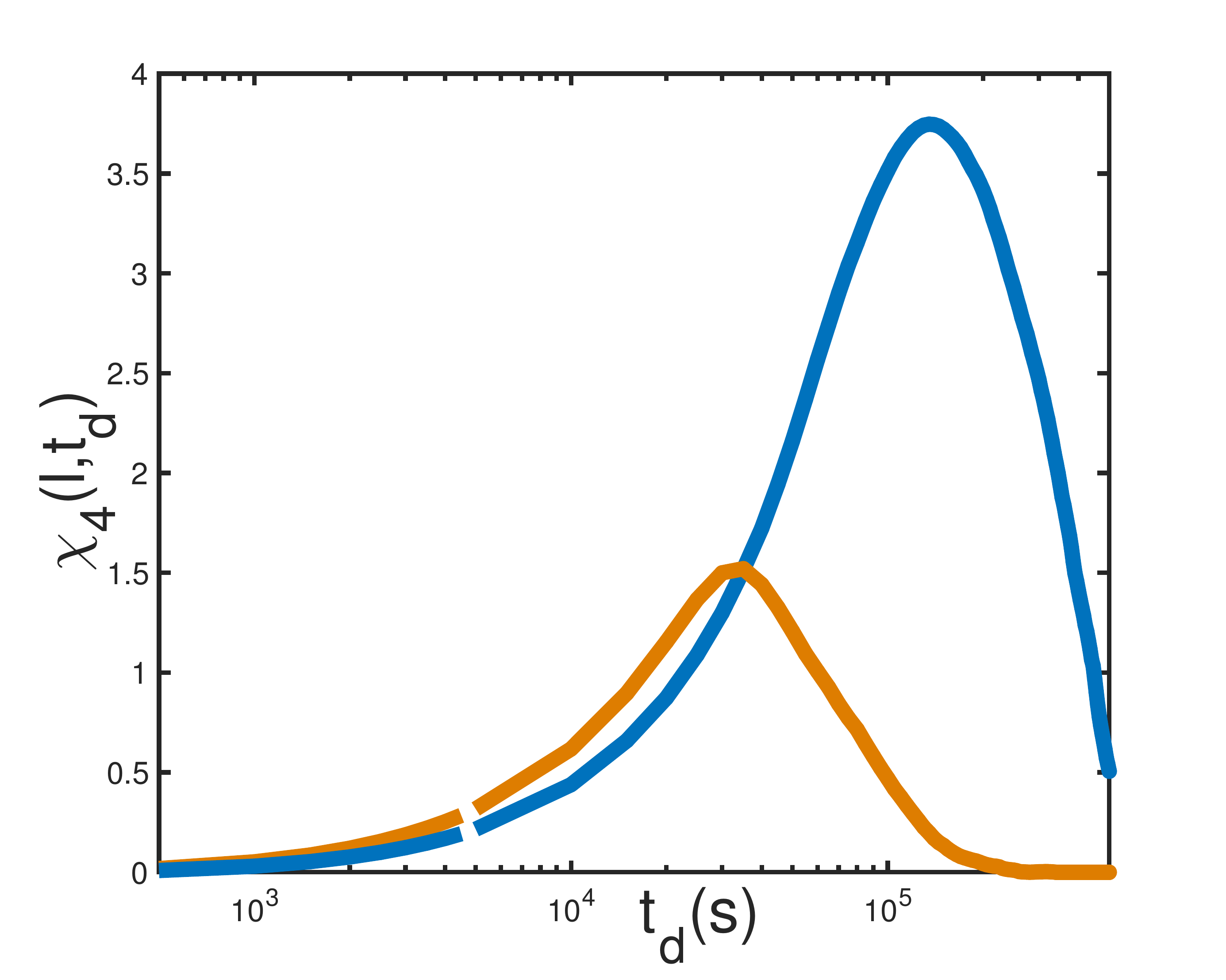} \label{xi_4_2_layer_sim} }
	\par
	\sidesubfloat[]{\includegraphics[width=0.55\textwidth] {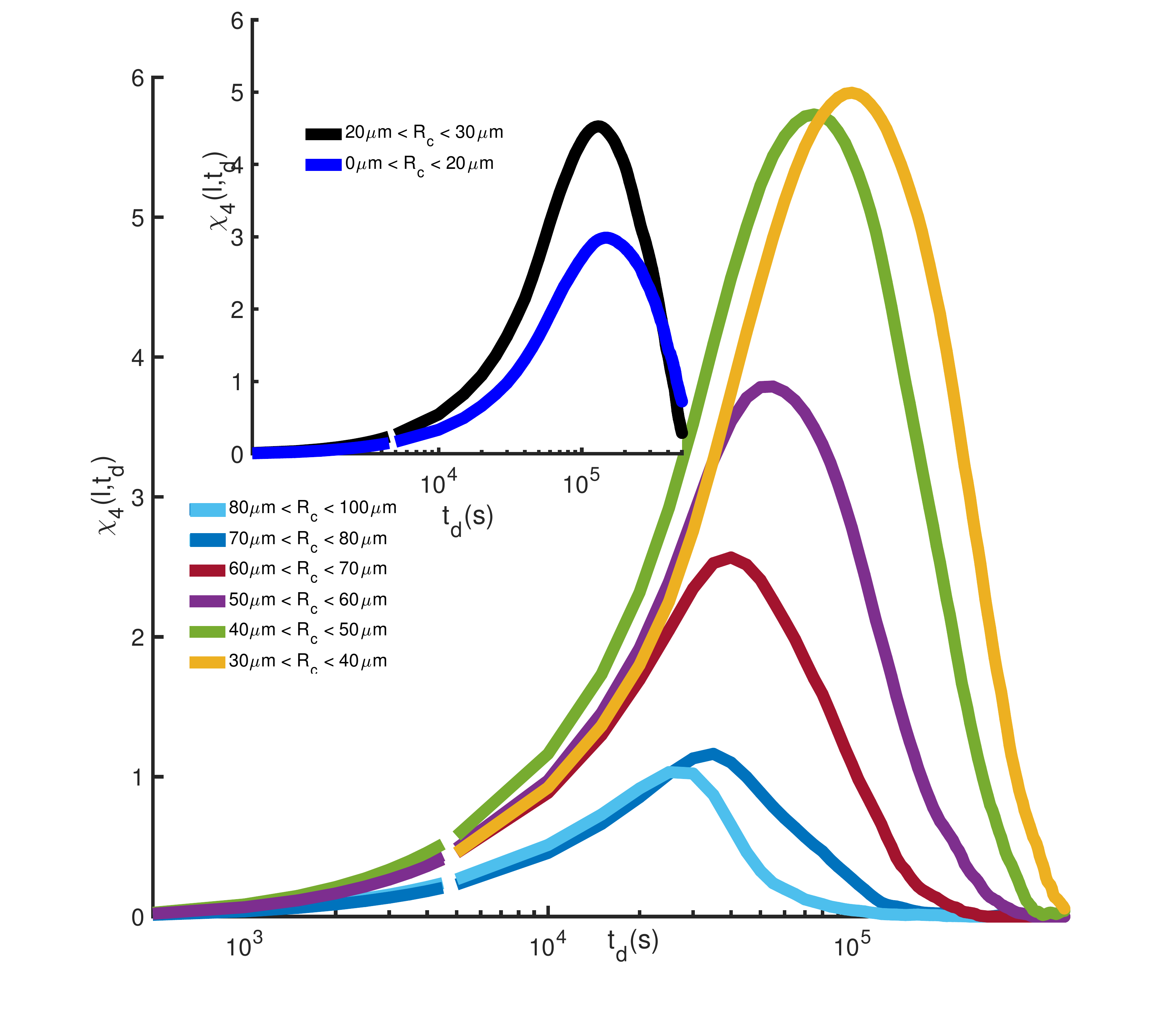} \label{chi4_sim} }
\caption{\textbf{(a)} Fourth Order Susceptibility ($\chi_4(l,t_d)$) determined by the variance in $\Omega(l,t_d)$ for experimentally tracked cells. $\chi_4(l,t_d)$ for cells in the core (periphery) is shown in blue (orange). \textbf{(b)} $\chi_4(l,t_d)$ for cells tracked in simulations. Blue (orange) line shows $\chi_4(l,t_d)$ for cells in the core (periphery).  \textbf{(c)} Layer by Layer Fourth Order Susceptibility ($\chi_4 (l,t_d)$) determined by variance in $\Omega(l,t_d)$. From top to bottom (except the innermost layer shown in sky blue) $\chi_4 (l,t_d)$ curves are for cells whose distance from the center of spheroid ($R_c$) are $10(i-1)~\mu m < R_{c} < 10i ~\mu m$, for all $i = \{4,....,8\}$. The innermost layer corresponds to  $80 \mu m < R_{c} < 100 \mu m$. In inset, black (blue) curve corresponds to $\chi_4(l, t_d)$ for  $20 \mu m < R_{c} < 30 \mu m$ ($0 \mu m < R_{c} < 20$).}
\label{xi_4_day_7}
\end{figure}

\clearpage
\floatsetup[figure]{style=plain,subcapbesideposition=top}
\begin{figure}
	\sidesubfloat[]{\includegraphics[width=0.55\textwidth] {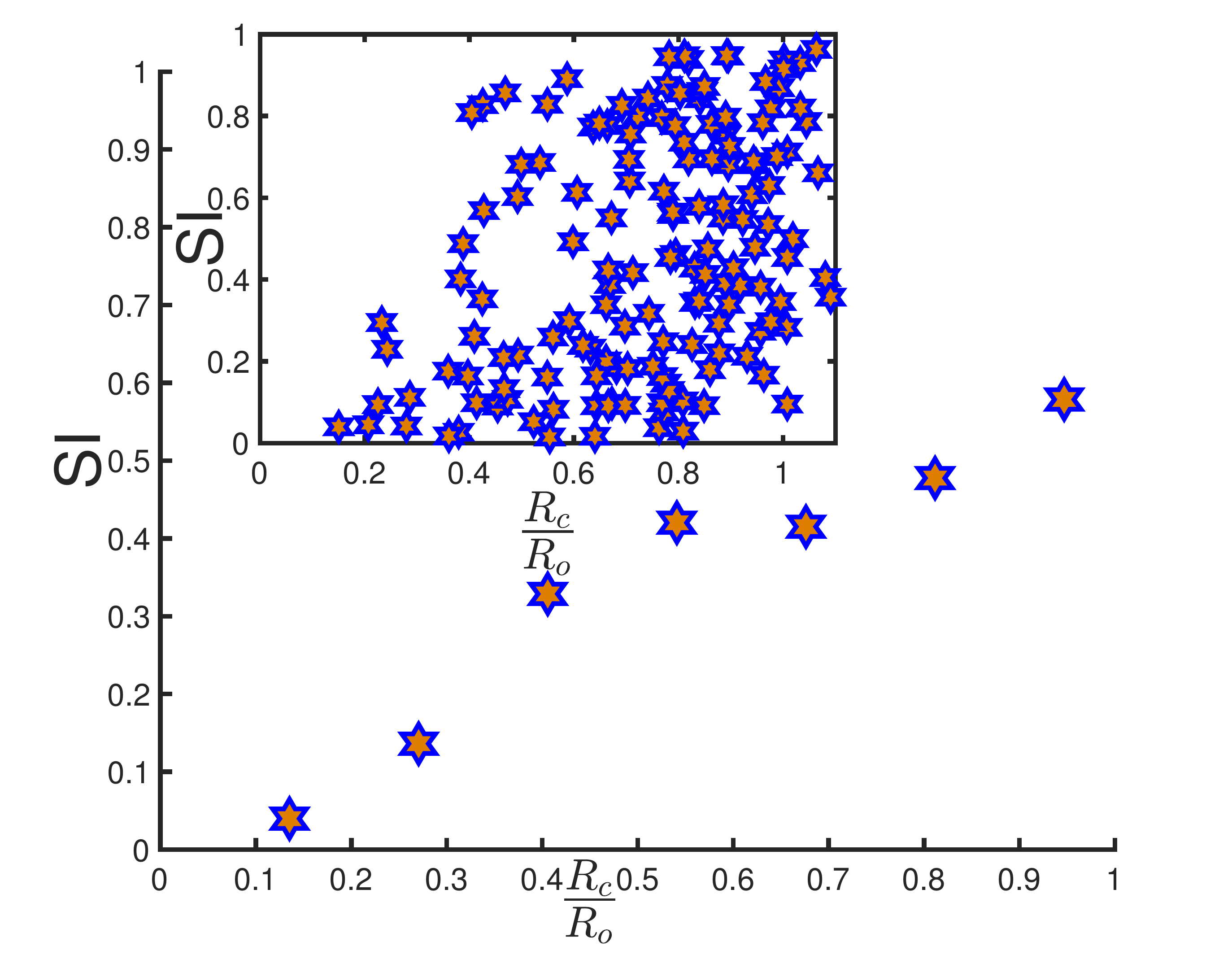} \label{straightness_expt} }
	\sidesubfloat[]{\includegraphics[width=0.52\textwidth] {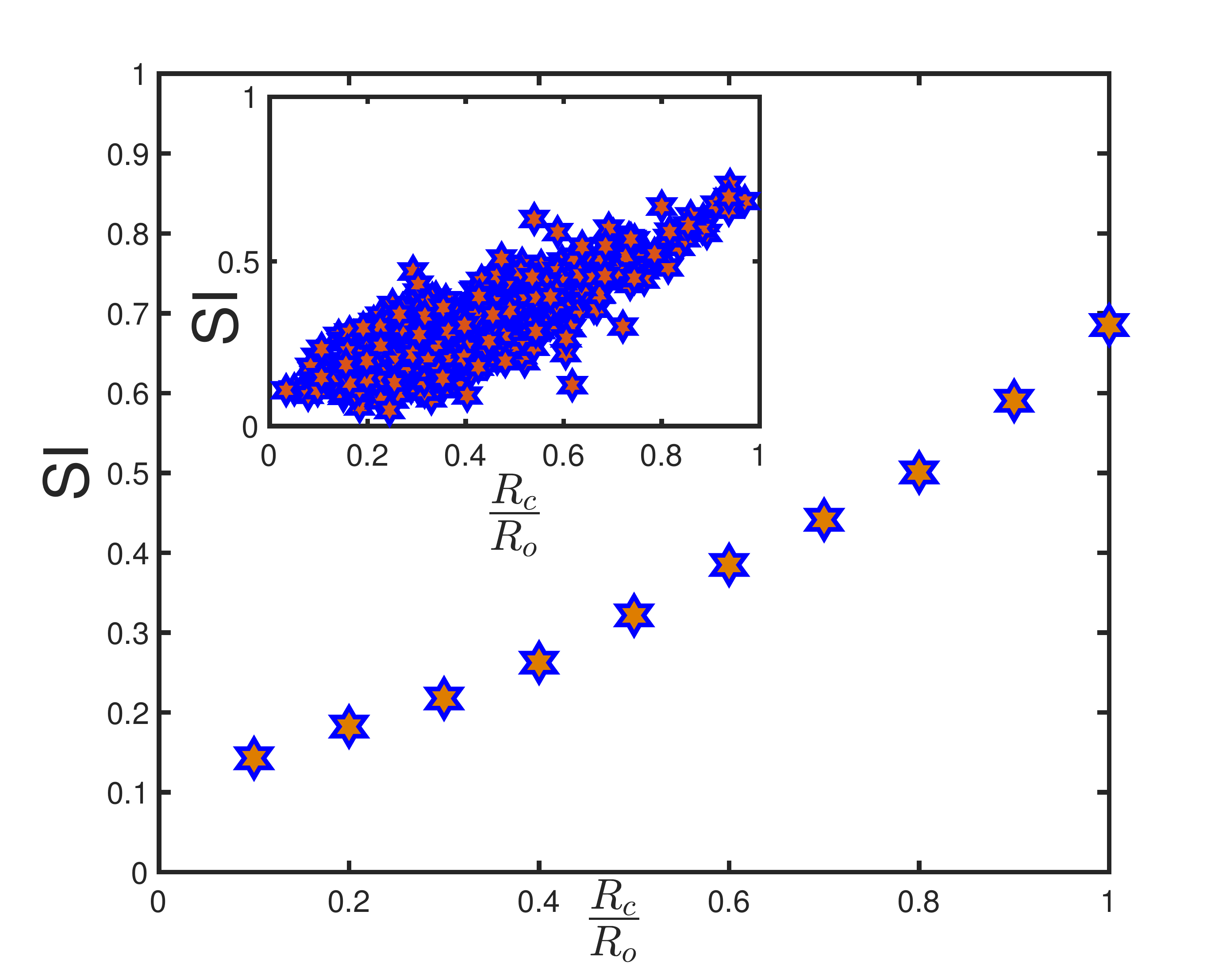} \label{straightness_sim} }
	\par
	\sidesubfloat[]{\includegraphics[width=0.53\textwidth] {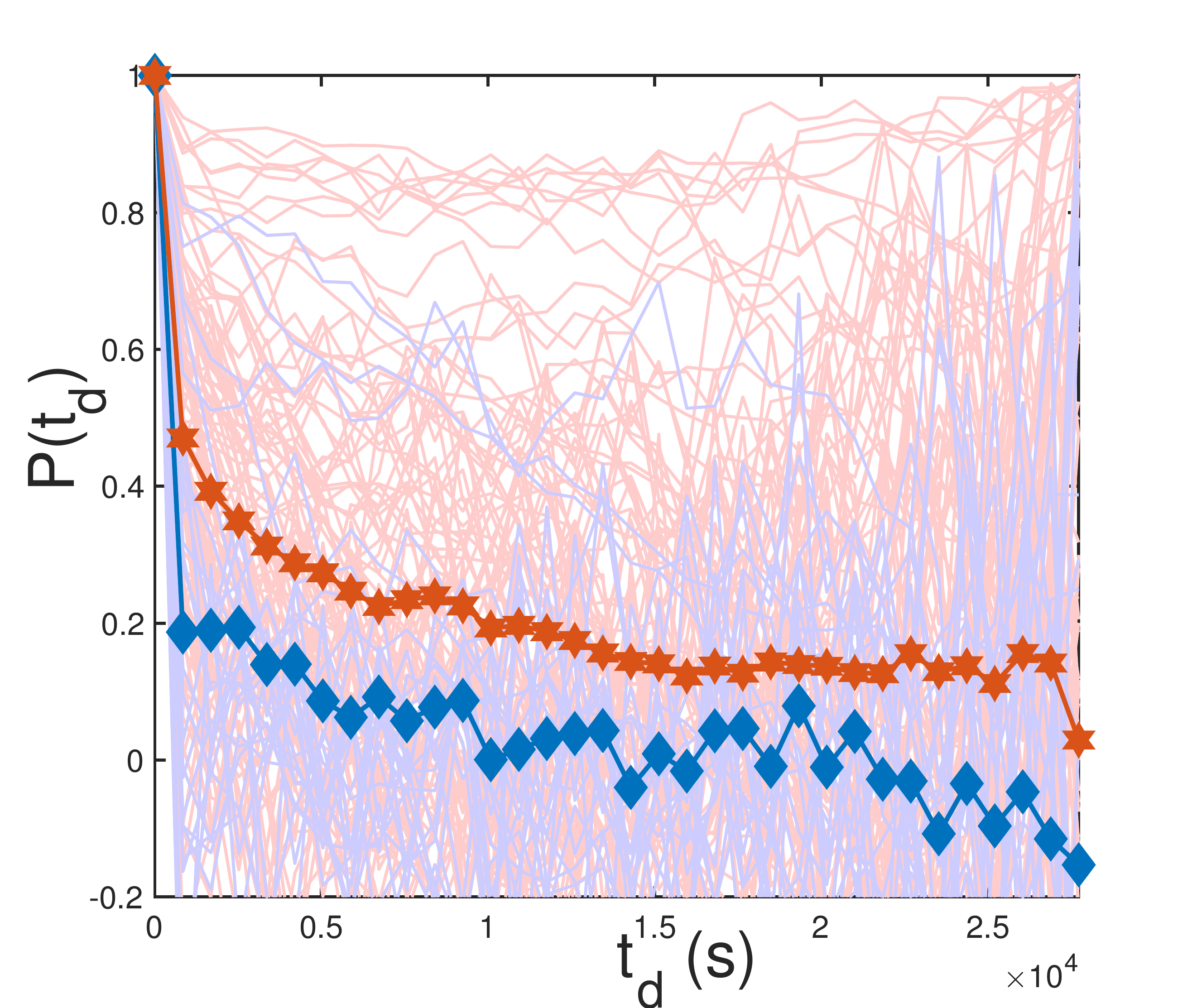} \label{vel_auto_expt} }
	\sidesubfloat[]{\includegraphics[width=0.57\textwidth] {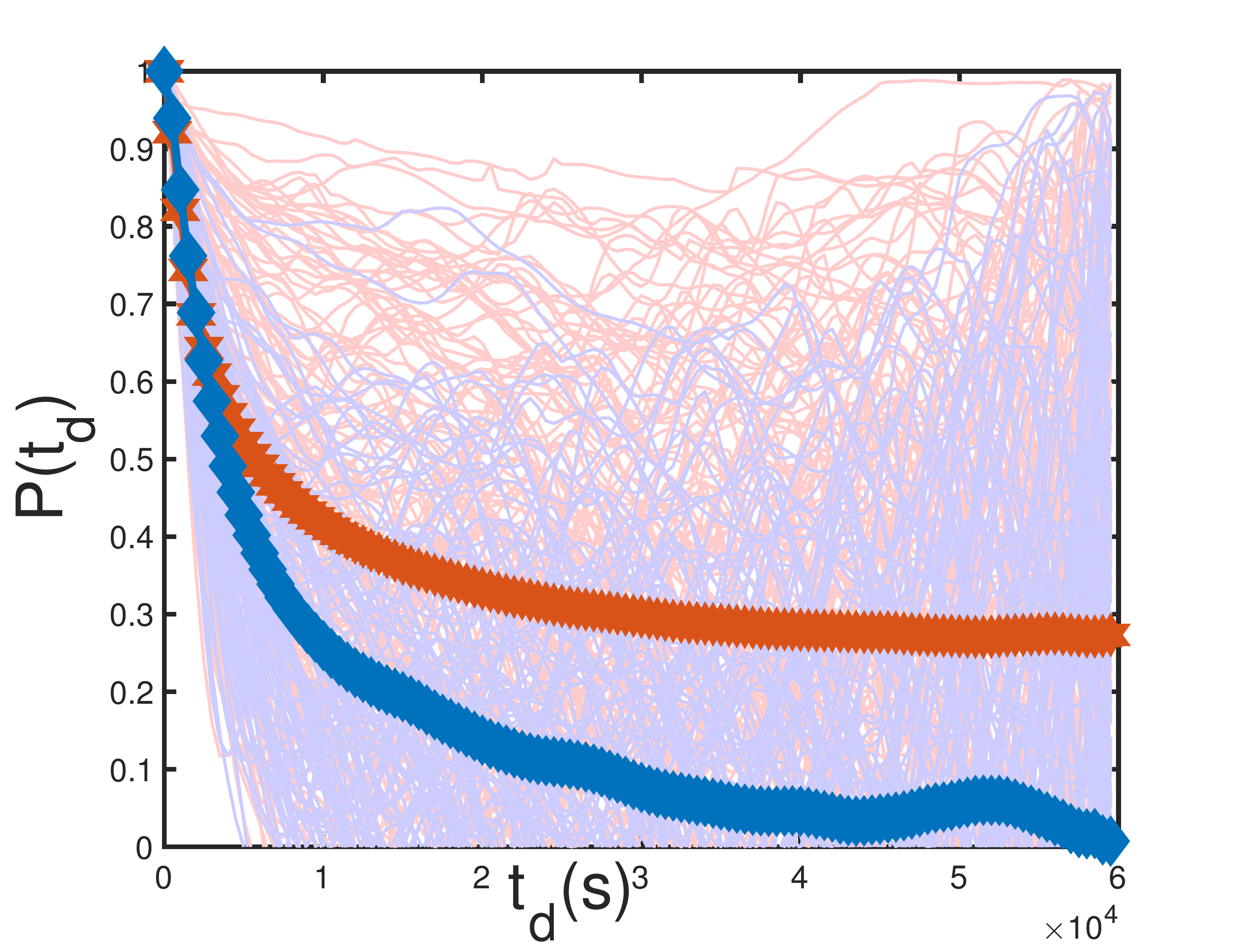} \label{vel_auto_sim} }
\caption{\textbf{(a)} Straightness Index ($SI$) of cells as a function of distance from the center of spheroid ($\frac{R_{c}}{R_o}$) obtained using the experimental data. The inset shows the scatter plot of $SI$ vs $\frac{R_{c}}{R_o}$ for all cells tracked. The plot in the main figure was generated by binning the data in the inset.  \textbf{(b)} SI for the simulated cells as a function of distance. The inset shows the SI for all the cells. The data in the inset was binned to generate the main figure. \textbf{(c)} Persistence ($P(t_d)$) function defined as $\langle \hat{v}(t+t_d)\cdot \hat{v}(t)\rangle_t$ for experimentally tracked cells. The red line depicts $P(t_d)$ for cells in the periphery ($R_{c} > 2~mm$) and blue line shows the $P(t_d)$ for cells in the core ($R_{c} < 1.5~mm$). The red and blue thin lines are $P(t_d)$ for individual cells. \textbf{(d)} $P(t_d)$ for simulated cells. The red (blue) line depicts $P(t_d)$ for cells in the periphery (core). The red (blue) thin lines are $P(t_d)$ for individual cells in periphery (core).}
\label{strightness_day_7}
 \end{figure}
 
 \clearpage
\begin{figure}
\contcaption{\textbf{(d)} $P(t_d)$ for simulated cells with red line for cells on the periphery ($R_{c} > 60 \mu m$) where as blue line is for cells in the core ($R_{c} < 30 \mu m$). The red and blue thin lines are $P(t_d)$ for individual cells. In both experiments and simulations there are substantial heterogeneities in individual cells.}
\end{figure}
\clearpage



\bibliography{main} 
\bibliographystyle{rsc} 

\end{document}